\documentclass[10pt,twocolumn,showpacs,longbibliography,amsmath,amssymb,osajnl,floatfix,superscriptaddress,asmart]{revtex4-2}

\usepackage{graphicx}
\usepackage{dcolumn}
\usepackage{bm}
\usepackage{color}
\usepackage{txfonts}
\usepackage{microtype}

\begin{document}

\title{Strong signature of one-loop self-energy in polarization resolved nonlinear Compton scattering}
\date{\today}

\author{Yan-Fei Li}\email{liyanfei@xjtu.edu.cn}	
\affiliation{Department of Nuclear Science and Technology, Xi'an Jiaotong University, Xi'an 710049, China}
\author{Yue-Yue Chen}\email{yueyuechen@shnu.edu.cn}
\affiliation{Department of Physics, Shanghai Normal University, Shanghai 200234, China}		
\author{K. Z. Hatsagortsyan}\thanks{k.hatsagortsyan@mpi-hd.mpg.de}
\author{A. Di Piazza}\email{New address: Department of Physics and Astronomy, University of Rochester, Rochester, NY (USA)
and Laboratory for Laser Energetics, University of Rochester, Rochester, NY (USA)}
\affiliation{Max-Planck-Institut f\"{u}r Kernphysik, Saupfercheckweg 1,
69117 Heidelberg, Germany}

\author{M. Tamburini}
\affiliation{Max-Planck-Institut f\"{u}r Kernphysik, Saupfercheckweg 1,
69117 Heidelberg, Germany}
\author{C. H. Keitel}
\affiliation{Max-Planck-Institut f\"{u}r Kernphysik, Saupfercheckweg 1,
69117 Heidelberg, Germany}
	
\begin{abstract}

The polarization dynamics of electrons including multiple nonlinear Compton scattering during the interaction of a circularly-polarized ultraintense laser pulse with a counterpropagating ultrarelativistic electron beam is investigated. While electron polarization emerges mostly due to spin-flips at photon emissions, there is  a non-radiative contribution to the polarization which stems from the one-loop QED radiative corrections to the self-energy,  
which admits of a simple physical model. We put forward a method to single out the non-radiative contribution to the polarization,
employing the reflection regime of the interaction when the radiation reaction is significant.
The polarization of electrons that penetrate in the forward direction through a colliding laser is shown to be dominated by the loop effect, while the reflected electrons are mostly polarized by spin-flips at photon emissions.  We confirm this effect by quantum Monte Carlo simulations considering the helicity transfer from the laser field to the electrons, taking into account the opposite sign of the polarizations induced by the non-radiative loop effect and radiative spin-flip. Our Monte Carlo simulations show a polarization signal as high as $\gtrsim 10\%$ from  the non-radiative effect,  amenable for  experimental detection with current technology.

\end{abstract}

\maketitle
\section{introduction}

 Ultrastrong laser technology is advancing rapidly \cite{Danson2015,Yoon2021}, which opens the door to laser-driven techniques for electron and ion acceleration, bright x-ray, and $\gamma$-ray generation \cite{Piazza2012,Corde2013,Gonoskov2022}.  In experiments, it rendered accessible nonlinear QED processes \cite{Ritus1985,Baier1998,bula1996observation,burke1997positron,bamber1999studies}, in particular, extreme multiphoton processes with radiation reaction \cite{cole2018experimental,poder2018experimental}, and electron-positron pair production \cite{Chen2010,Sarri2013}, and will offer a test bed for the nonperturbative QED \cite{ritus1970radiative,yakimenko2019prospect}.

A recent new twist in the theory of nonlinear QED is the scrutiny of possibilities to employ ultrastrong laser fields for ultrafast polarization of electrons \cite{DelSorbo2017,Seipt2018,Liyf2019,Seipt2019,li2022helicity} and positrons \cite{Chen2019,Liyfei2020,Wan2020} in a femtosecond timescale. While radiative electron polarization (RP), has been known from early works on synchrotron radiation (Sokolov-Ternov effect) \cite{Sokolov1964,baier1971radiative,Baier1972,Derbenev_1973}, its realization with laser fields has been hindered by the oscillating symmetric character of the laser magnetic field \cite{kotkin1998polarization,Kotkin2003,Ivanov_2004,Ivanov_2005} such that specific setups are necessary \cite{Liyf2019,Seipt2019,Chen2019,Liyfei2020,li2022helicity} to break the symmetry and to yield sizable polarization.

The main reason for the electron RP is the spin-flip during photon emission. However, even if all emissions occur without spin-flip, polarization may emerge because of the dependence of the photon emission probability on the electron's initial spin  projection on the magnetic field  in the instantaneous rest frame of the particle. For instance, in an initially unpolarized beam, the electrons that radiate will be polarized even without spin-flip because the emission is preferred in a certain spin state. Accordingly, the electrons in the beam that do not emit will be polarized oppositely to the radiating ones because, in the absence of spin-flip, the initial unpolarized state of the total beam cannot change \cite{Kotkin2003}. In this case the polarization can arise not in the whole beam but only in the separated parts of the radiating and non-radiating electrons.
The latter provides a simple explanation (simple man's model) for the electron non-radiative polarization (NRP), as well as for the, at first sight unexpected, probability of the electron polarization change via the ``no-photon-emission" process \cite{CAIN}. 
In other words, if the radiating and not radiating electrons are mixed in a beam, its polarization can arise only due to the additional effect of the spin-flip. Nevertheless, the NRP effect (i.e., no-photon-emission effect connected with the QED loop diagram, see the discussion below on Fig. 1)  may become observable with a proper selection of electrons after the interaction, which will be the main aim of this paper.

While the simple man's model helps for an intuitive understanding, a quantum description of NRP within nonlinear QED is in the order. In strong background fields, the electron polarization dynamics in the case of a small value of the quantum strong-field parameter $\chi \equiv |e| \sqrt{-(F_{\mu \nu}p^v)^2}/m^3\ll 1$ has been described by Baier using the operator approach of strong field QED \cite{Baier1972}. Here $F_{\mu \nu}$ denotes the field tensor,  $p^\nu$ the 4-momentum of the electron, $e<0$ and $m$ the electron charge and mass, respectively. The relativistic units $\hbar=c=1$ are used throughout. The expectation value of the spin is calculated, expanding the time evolution operator up to the second order over the coupling with the radiation photon field. The first-order term describes the photon emission effect and is related to the polarization effects due to photon emission. The second-order term stems from the non-radiative one-loop self-energy (OLSE) interaction.  While the real part of the OLSE contribution describes spin rotation 
 associated to the anomalous magnetic moment, additional to the common spin precession due to the electron's Dirac magnetic moment, the imaginary part of OLSE (related according to the optical 
theorem to the photon emission probability) introduces 
 spin-dependent damping 
terms for polarization, which correspond to the NRP effect. The combination of all damping terms in Ref. \cite{Baier1972} yields solely the probability of the polarization due to the radiative spin-flip in consistence with the simple man's model presented in Sec. \ref{simpleman}, which tells us that the NRP and RP effects without spin-flip cancel each other if no selection of radiating (non-radiating) electrons is implemented. The impact of the OLSE correction on the electron spin dynamics in intense background fields has been demonstrated also in \cite{Meuren2011,podszus2021first} solving the Dirac equation with the QED mass operator (Schwinger-Dirac equation), and in \cite{torgrimsson2021resummation,torgrimsson2021resummation}, via the resummation method using M\"uller matrices.

\begin{figure}
 	\includegraphics[width=0.45\textwidth]{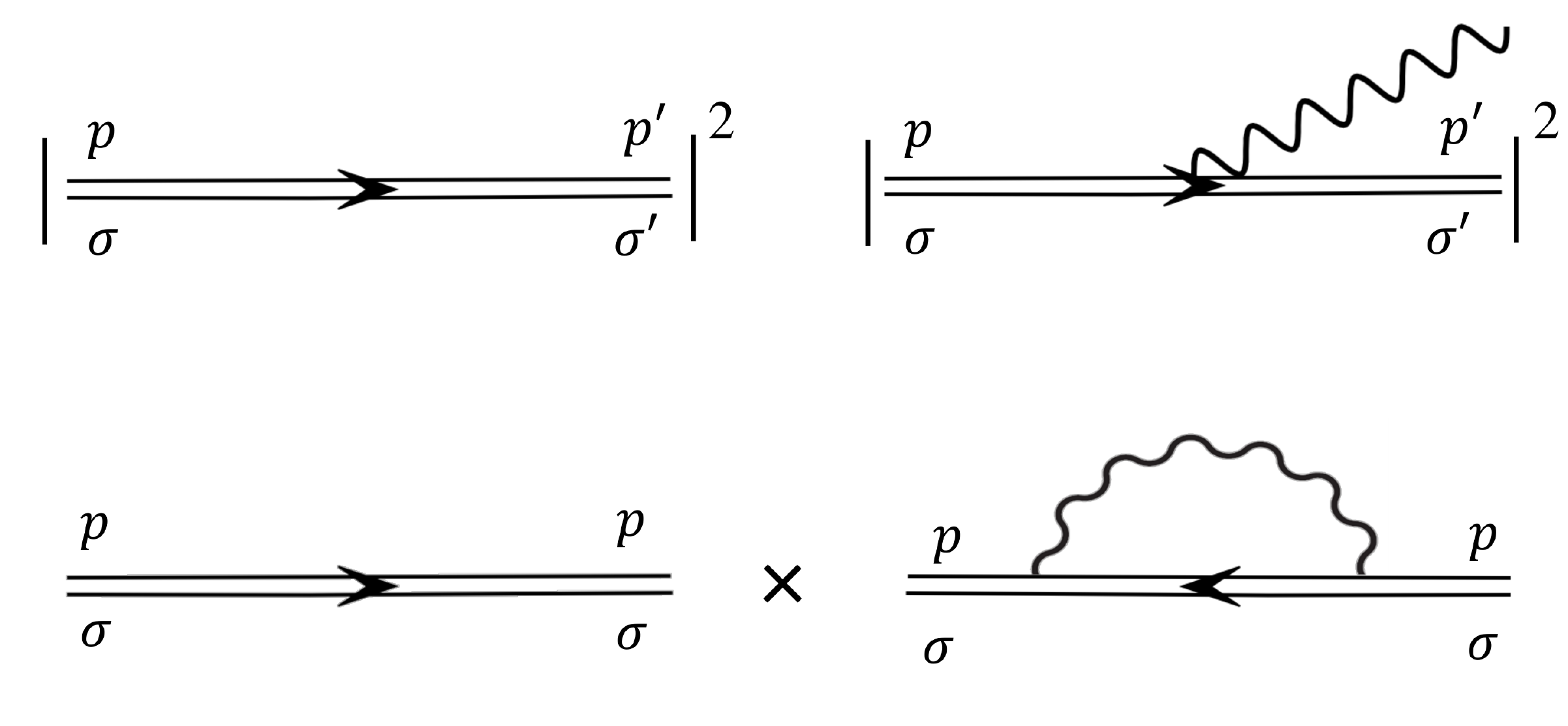}
  \begin{picture}(300,-30)
  \put(5,105){\small (a)}
  \put(120,105){\small(b)}
  \put(5,45){\small (c)}
 	  \end{picture}
  	  \caption{Feynman diagrams up to the first order of $\alpha$ contributing to the polarization of an electron in a background strong field: (a) spin precession in a background field (corresponding to the BMT equation);  (b) radiative polarization due to the spin-flip and the spin dependence of the photon emission probability; (c) the OLSE contribution, including NRP,
  and the modification of the spin precession due to the anomalous magnetic moment. Here, $\rho (\sigma)$ and $\rho' (\sigma')$ are the initial and final momentum (spin).}
\label{diagrams}
\end{figure}
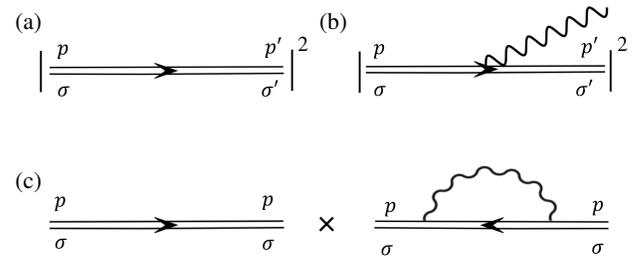

The polarization picture given by Baier can be illustrated using the QED technique of Feynman diagrams in Furry representation \cite{Ilderton2020}, which is also valid at high $\chi$, see  Fig.~\ref{diagrams}. The first diagram $\sim \alpha^0$ (with the fine structure constant $\alpha$) is related to the electron spin precession known from the Bargmann-Michel-Telegdi (BMT) equation \cite{Bargmann1959}, and the second one ($\sim \alpha$) describes the RP at a photon emission, which includes two contributions, one due to the spin-flip, and one due to the spin dependence of the photon emission probability. The NRP is described by the third, interference diagram ($\sim \alpha$) of the OLSE diagram with the forward scattered one \cite{Baier1972,Torgrimsson_2021}.
It includes NRP due to the spin dependence of the photon emission probability, and the modification of the spin precession due to the anomalous magnetic moment.
 Note that the simple man's model and QED calculations yield the same probability for the NRP, expressed via the spin-dependent radiation probability.

Thus, in physical terms, the leading contribution to electron polarization in a strong background field of the first order in $\alpha$ is induced by the following three processes: the spin-flip during photon emission, the spin dependence of the photon emission probability (which induces NRP, but has a contribution for radiating electrons as well), and the spin precession modification due to the electron anomalous magnetic moment. Note the spin precession via BMT equation does not yield electron polarization because the relative pointing directions of the electron velocity and spin remain unchanged \cite{Meuren2011,Ilderton2020}. The spin-flip is employed for electron beam polarization in Refs.~\cite{Liyf2019,Seipt2019,li2022helicity}. We have also discussed the possibility of the observation of the electron polarization caused by the electron anomalous magnetic moment~\cite{li2022helicity}.

The NRP is described by the OLSE. To observe the distinct signature of the OLSE in the electron polarization, one needs to face the challenge of the separation of the electrons experiencing substantial  radiation recoil from those with negligible recoil in the beam, because the polarization due to radiation and NRP are both of the same order of magnitude ($\sim \alpha$) and mixed for the total beam. The OLSE polarization effect was discussed in Refs.~\cite{Meuren2011,Ilderton2020,Torgrimsson_2021}.
Ref.~\cite{Meuren2011} shows that the electron's exact spin-dependent wave function is unstable inside a linearly polarized laser field, leading to $\sim 1\%$ longitudinal polarization due to OLSE at $\chi\sim 1$, being rotated from the initial transverse polarization (100\%) for unscattered electrons, $\sim 10^{-4}$ out of the total amount (in a linearly polarized laser field, the net polarization is negligible due to the averaging of polarization effects in oscillating fields). The use of an ultrashort laser pulse allows one in this scheme to neglect the amount of radiating electrons. In Ref.~\cite{Torgrimsson_2021}, deriving the spin-resolved probabilities for nonlinear Compton scattering, 
 the electron polarization via OLSE is discussed, however, without a specific setup and estimating the experimental feasibility. Another possibility for the detection of the OLSE polarization effect was proposed in Ref.~\cite{Ilderton2020}, working in the $\chi\ll 1$ regime, when the spin-flip at photon emission is damped more strongly than the OLSE one. The obtained OLSE polarization signature in \cite{Ilderton2020} was of $\alpha^2$-order with a suppressed probability $w\leq 10^{-6}$. Unfortunately, the signal predicted in these schemes is far below the current experimental detection precision, which is typically $\gtrsim 0.5\%$ \cite{Narayan2016}.

\begin{figure}[b]
 	\includegraphics[width=0.4\textwidth]{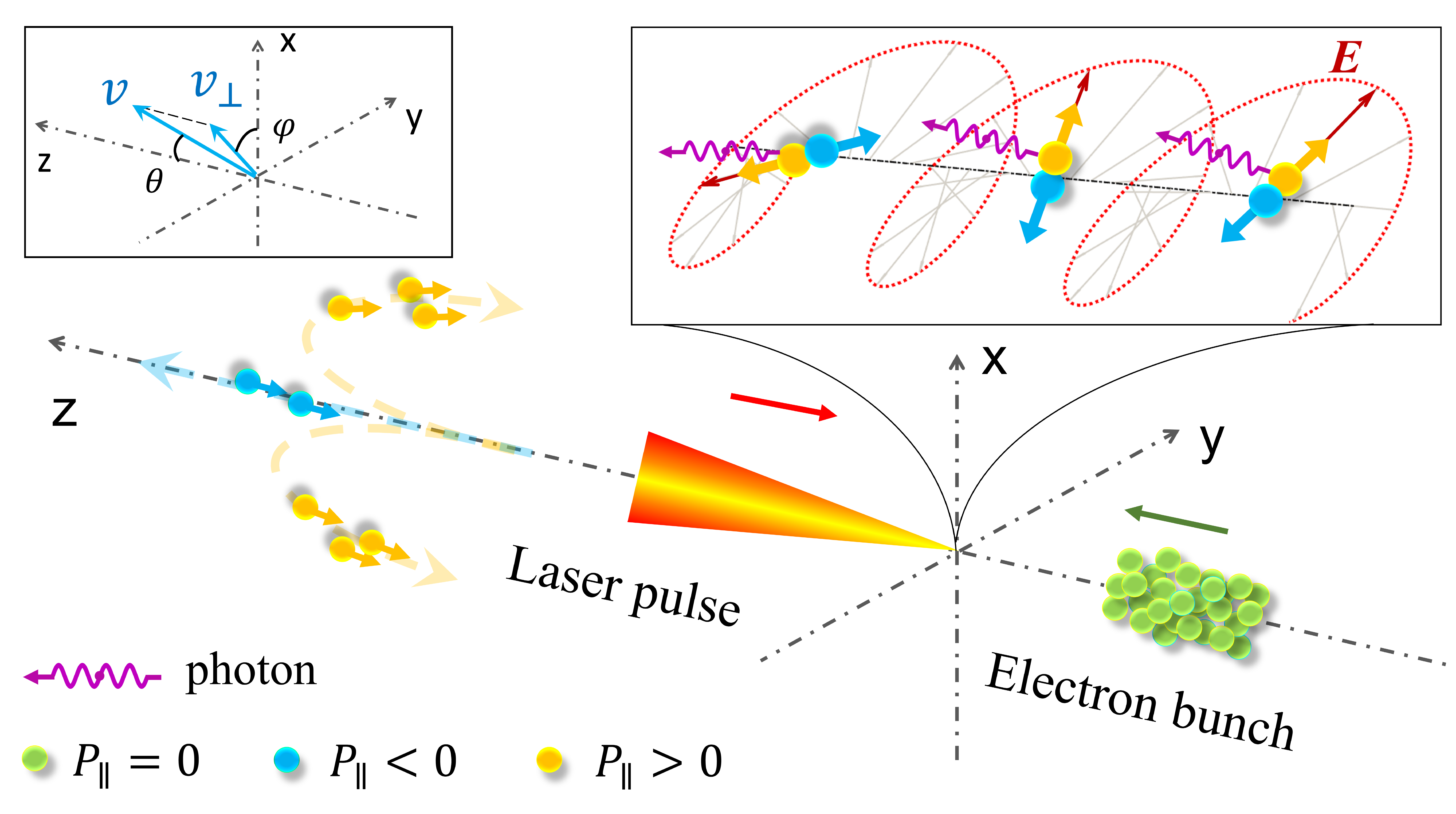}
 	  \caption{The scheme for the detection of OLSE during nonlinear Compton scattering of a CP strong laser pulse by an unpolarized electron beam. The instantaneous polarization caused by RP (NRP)  is rotated into a longitudinal direction through spin precession. After the interaction, the electrons dominated by  RP (NRP) have a final longitudinal polarization with $P_\parallel>0$ ($P_\parallel<0$). While the electrons undergoing a substantial radiative 
 	   recoil have $P_\parallel>0$ and are reflected from the laser pulse, the electrons moving forward have $P_\parallel<0$ and indicate the signature of OLSE. }
\label{fig1}
\end{figure}

In this paper, we put forward a method for the observation of the OLSE effect of the first order in $\alpha$, in the polarization dynamics of electrons during the interaction with a circularly-polarized (CP) ultraintense laser pulse in the quantum radiation-dominated regime with $\chi\sim 1$.
To this end, the separation of  electrons experiencing a weak radiation recoil from those with  significant recoil is realized by employing the quantum reflection regime, when deflection of electrons to large angles arises because of radiation reaction \cite{Dipiazza2009,Li2015,Geng2019}. The spin-polarization of the deflected electrons is dominated by the RP, while that of the forward-moving ones (at small deflection angles) is dominated by the NRP because of small energy loss. We demonstrate that there exists a small angle domain in which NRP dominates over RP, providing a distinct signature for the NRP effect, see Fig. \ref{fig1}. While the longitudinal polarization of electrons coincides with the driving-laser helicity as RP dominates, it is the opposite when NRP dominates. An important point is that we employ a polarization mechanism that is based on the anomalous magnetic moment of the electron induced by OLSE. In this mechanism the helicity of the laser photons  is transferred to the electron helicity, enabled by the spin precession via the $g - 2$ term of the BMT equation. In this process, the phase matching of the transverse electron polarization and the laser field leads to the accumulation of helicity through the oscillating laser field and to the enhancement of the OLSE signal. 

The structure of the paper is the following. In Sec.~\ref{simpleman} we will introduce a simple man's model for the intuitive explanation of the NRP effect, which is augmented with the known QED description  in Sec.~\ref{QED}. The  detection scheme of M{\o}ller polarimetry and results of the numerical QED Monte Carlo simulations are presented in Sec. ~\ref{result}.  The experimental feasibility is analyzed in Sec.~ \ref{experiment}. The impact of laser and electron beam parameters on the OLSE signature, efficiency of the detection with energy-selection technique, and the influence of pair-production effect are discussed. Our conclusion is given in Sec.~\ref{conclusion}.


\section{layperson’s model}\label{simpleman}

Following the tradition of the strong-field atomic physics \cite{Corkum_1993,simpleman}, in this section, we provide a simple man's model similar to that in Ref. [33] describing the emergence of the electron beam polarization during photon emissions in a strong background field. 

\subsection{No spin-flip}

 In this sub-section we discuss the electron beam polarization which arises due to the spin dependence of the photon emission probability. The latter has a consequence not only for the polarization of the radiating electrons, but also for the polarization of the non-radiating part of the beam.  In the next sub-section we extend the simple man's model including the spin-flip effect during a photon emission.

Let us assume that the electron beam consisting of $N$ electrons is initially partially polarized with the average spin projection $S_i$ on the given quantization axis  along the  magnetic field of the laser (for the sake of convenience, here and below we often indicate as ``spin'' quantities which are actually twice larger):
\begin{eqnarray}
S_{i}=\frac{N^+-N^-}{N },\label{si}
\end{eqnarray}
where $N^\pm$ are the number of fully polarized electrons in the initial beam
with positive and negative spin projections, and
\begin{eqnarray}
N&=&N^++N^-.\label{NNN}
\end{eqnarray}
We can find from Eqs.(\ref{si})-(\ref{NNN}) that
\begin{eqnarray}
\frac{N^\pm}{N}=\frac{1\pm S_i}{2}.\label{ssi}
\end{eqnarray}
Let us denote the number of electron in the beam that emit photons by $N_R$, and that with no-photon emission $N_{NR}$,
\begin{eqnarray}
N&=&N_{R}+N_{NR}.
\end{eqnarray}
If $N_R^\pm$ are the number of radiating electrons in the initial beam with the positive and negative spin projections, and $N_{NR}^\pm$ that of the non-radiating electrons:
\begin{eqnarray}
N^\pm&=&N_{R}^\pm+N_{NR}^\pm \label{N+-}\\
N_R&=&N_R^++N_R^-\\
N_{NR}&=&N_{NR}^++N_{NR}^-
\end{eqnarray}
Firstly, assume that no spin-flip takes place, but the photon emission probability depends on the electron initial spin, which can be expressed as
\begin{eqnarray}
w_R^\pm=\frac{N_R^\pm}{N^\pm}=w_0\pm w_s,\label{rad}
 \end{eqnarray}
  which follows from the the photon emission probability in  locally-constant-field approximation (LCFA) in the $u$-interval $\Delta u$ and time-interval $\Delta t$ summed up over the final polarization \cite{Baier1998}:
\begin{eqnarray}
\label{W1}
 w_R  &=&\Delta u\Delta t C_R(a_{1}+{\bf a}_{2}\cdot{\bf S}_{i} ),
\end{eqnarray}
with \begin{eqnarray}a_1&=&-(1+u){\rm IntK}_{\frac{1}{3}}(u')+(u^2+2u+2){\rm K}_{\frac{2}{3}}(u'),\nonumber\\{\bf a}_2&=&-\hat{{\bf e}}_2 u {\rm K}_{\frac{1}{3}}(u'),\nonumber\\C_R&=&\frac{\alpha m}{{\sqrt{3}\pi \gamma_e}{\left(1+u\right)^3}},\nonumber\end{eqnarray}where $\gamma_e$ the electron Lorentz factor, $u=\omega_\gamma/\left(\varepsilon_i-\omega_\gamma\right)$,  $\omega_\gamma$ the emitted photon energy, $\varepsilon_i$ the electron energy before radiation, $u'=2u/3\chi_e$, ${\rm IntK}_{\frac{1}{3}}(u')\equiv \int_{u'}^{\infty} {\rm d}z {\rm K}_{\frac{1}{3}}(z)$,  ${\rm K}_n$ the $n^{th}$-order modified Bessel function of the second kind, $\hat{{\bf e}}_1$ the unit vector along the  transverse component of the electron acceleration, $\hat{{\bf e}}_2=\hat{\bf e}_v\times\hat{\bf e}_1$,  $\hat{\bf e}_v$ the unit vector along the electron velocity, ${\bf S}_{i}$ the electron spin vector before radiation.
Thus, 
\begin{eqnarray}\label{NR2}
w_0&=& \Delta u\Delta t C_R a_{1}; \,\,\,\,\,w_s= \Delta u\Delta t C_R  a_{2},
\end{eqnarray}
where the quantization axis is assumed along $-\hat{{\bf e}}_2$, and $a_{2}=|\mathbf{a}_2|$.

The polarization of the part of the beam of radiating electrons due to the spin dependence of the photon emission probability
is
\begin{eqnarray}
S_R^{\rm no-sf} &=& \frac{N_{R}^+-N_{R}^-}{N_{R} }= \frac{\frac{N_R^+}{N^+}\frac{N^+}{N }-\frac{N_R^-}{N^-}\frac{N^-}{N }}{\frac{N_R^+}{N^+}\frac{N^+}{N }+\frac{N_R^-}{N^-}\frac{N^-}{N }}\nonumber\\
 &=& \frac{(w_0+w_s)\frac{1+S_i}{2}-(w_0-w_s)\frac{1-S_i}{2}}{(w_0+w_s)\frac{1+S_i}{2}+(w_0-w_s)\frac{1-S_i}{2}}\nonumber\\
 &=&\frac{w_s+S_iw_0}{w_0+S_iw_s},\label{sr1}
\end{eqnarray}
 where we have used Eqs.~(\ref{ssi}) and (\ref{rad}) and the ``no-sf" superscript indicates that the spin-flip is not accounted for in these expressions. .

The polarization of the part of the beam of non-radiating electrons due to the spin dependence of the photon emission probability, i.e. NRP,  is
\begin{eqnarray}
S_{NR}=\frac{N_{NR}^+-N_{NR}^-}{N_{NR} }=\frac{S_iN-(N_R^+-N_R^-)}{N-N_{R}},
\end{eqnarray}
where we have used Eqs.~(\ref{si}) and (\ref{N+-}). Then, we have
\begin{eqnarray}
S_{NR}&=&\frac{S_i-\left(\frac{N_R^+}{N^+}\frac{N^+}{N }-\frac{N_R^-}{N^-}\frac{N^-}{N }\right)}{1- \left(\frac{N_R^+}{N^+}\frac{N^+}{N }+\frac{N_R^-}{N^-}\frac{N^-}{N }\right)}\nonumber\\
&=& \frac{S_i-(w_s+S_iw_0)}{1-(w_0+S_i w_s)}\equiv \frac{B}{A},\label{snr}
\end{eqnarray}
again via Eqs.~(\ref{ssi}) and (\ref{rad}).

Thus, we conclude that non-radiating part of the electron beam can be also polarized because of the spin dependence of the radiation probability. The change of the polarization of non-radiating electrons is
\begin{eqnarray}
S_{NR}-S_i&=& \frac{(S_i^2-1)w_s}{1-(w_0+S_iw_s)} ,\label{snr2}
\end{eqnarray}
i.e. it is  vanishing only for the initial beam fully polarized along the magnetic field of the laser, $S_i^2=1$.


The total beam polarization solely due to the spin dependence of the photon emission probability is naturally not changing after the interaction if no selection of radiating (non-radiating) electrons is  carried out:
\begin{eqnarray}
 S_{\rm tot}^{\rm no-sf}= \frac{N_R S_R^{\rm no-sf}+N_{NR}S_{NR}}{N}=\frac{N^+-N^-}{N}=S_i.
\end{eqnarray}
In the most simple case of the initially unpolarized electron beam $S_i=0$:
\begin{eqnarray}
S_R^{\rm no-sf}=\frac{w_s}{w_0}\,\,\,\,\,\,\,\,\,
S_{NR}=-\frac{w_s}{1-w_0},
\end{eqnarray}
with $N_R/N=w_0$ and $N_{NR}/N=1-w_0$, and the total beam polarization is vanishing $S_{\rm  tot}^{\rm no-sf}=0$. Nevertheless, if radiating electrons will be separated from the nonradiating ones both sub-beams will be polarized even without accounting for the spin-flip during the photon emission.

The equation~(\ref{snr}) gives the average spin of the non-radiating electron, from which  the NRP probability is recovered
\begin{eqnarray}
W_{NR}&=& \frac{1}{2}(A+S_fB)\nonumber\\
&=&\frac{1}{2}\left\{1-(w_0+S_iw_S)+S_f \left[S_i-(w_s+S_iw_0)\right]\right\}.\label{npep}
\end{eqnarray}
While Eq.~(\ref{snr}) provides the ratio $B/A$, the prefactor in Eq.~(\ref{npep}) is chosen to yield for the spin-averaged total probability $W_{NR}=1-w_0$.

Thus, with the spin dependent radiation probability in the form of Eq.~(\ref{rad}), one can derive NRP probability Eq.~(\ref{npep}), 
 which taking into account Eqs.~(\ref{W1}), (\ref{NR2}), will read:
\begin{eqnarray}\label{NR3}
 W_{NR}  &=&\frac{1}{2}\left\{1-\Delta u\Delta t C_R (a_{1}+{\bf a}_{2}\cdot{\bf S}_{i} )\right.\nonumber\\
 &+&\left.{\bf S}_{f}\cdot\left[{\bf S}_{i}\left(1- \Delta u\Delta t C_R a_1 \right)-\Delta u\Delta t C_R\,{\bf a}_{2}\right]\right\}.
\end{eqnarray}
 The same NRP probability is derived within the nonlinear QED theory from the OLSE diagram, see Sec.~\ref{OLSE_calc} below. 

Note that Eqs.~(\ref{sr1}) and (\ref{snr2}) for the polarization change due to the spin dependence of the radiation probability can be expressed, using expressions of Eq.~(\ref{NR2}), as
\begin{eqnarray}
 \mathbf{S}_R^{\rm no-sf}-\mathbf{S}_i&=&\frac{\mathbf{a}_2-\mathbf{S}_i (\mathbf{a}_2 \cdot \mathbf{S}_i)}{a_1+\mathbf{S}_i \cdot \mathbf{a_2}}\\
 \mathbf{S}_{NR}-\mathbf{S}_i&=& \frac{\mathbf{S}_i (\mathbf{a}_2 \cdot \mathbf{S}_i) -\mathbf{a}_2}{1-(a_1+\mathbf{S}_i \cdot \mathbf{a_2})\Delta u\Delta t C_R }\Delta u\Delta t C_R.\nonumber\\
\end{eqnarray}
The average spin changes for the radiating and non-radiating electrons, without accounting spin-flip, are:
\begin{eqnarray}
 \Delta \mathbf{S}_R^{\rm no-sf} &=& (\mathbf{S}_R^{\rm no-sf}-\mathbf{S}_i)\frac{N_R}{N}\nonumber\\
 &=&-[\mathbf{S}_i (\mathbf{a}_2 \cdot \mathbf{S}_i)-\mathbf{a}_2]\Delta u\Delta t C_R \label{DSF}\\
 \Delta \mathbf{S}_{NR} &=& (\mathbf{S}_{NR}-\mathbf{S}_i)\frac{N_{NR}}{N}\nonumber\\
 &=&[\mathbf{S}_i (\mathbf{a}_2 \cdot \mathbf{S}_i) -\mathbf{a}_2]\Delta u\Delta t C_R .\label{DSNR}
\end{eqnarray}
The latter shows that without spin-flip the total beam polarization does not change $ \Delta \mathbf{S}_R^{\rm no-sf}+ \Delta \mathbf{S}_{NR}=0$.

\subsection{The electron beam polarization including the spin-flip effect}

Now let us consider the spin-flip effect during a photon emission for the electron beam polarization. Assume the electron spin-flip can happen due to the photon emission with a probability $w_{\rm sf}^\pm$, and it depends on the initial spin state. Then, after the spin-flip we have for the number of radiating electrons:
\begin{eqnarray}
N_R^{\rm sf+}=N_R^- w_{\rm sf}^-+(1-w_{\rm sf}^+)N_R^+\\
N_R^{\rm sf-}=N_R^+ w_{\rm sf}^++(1-w_{\rm sf}^-)N_R^-.
\end{eqnarray}
Of course the total number of radiating electrons does not change because of the spin-flip $N_{Rf}=N_R^{\rm sf+}+N_R^{\rm sf-}=N_R$, and
\begin{eqnarray}
N_R^{\rm sf+}-N_R^{\rm sf-}=(1-2w_{\rm sf}^+)N_{R}^+-(1-2w_{\rm sf}^-)N_R^- .
\end{eqnarray}

With the spin-flip, the polarization of the radiating electrons will read:
\begin{eqnarray}
S_{R}^{\rm sf} &=& \frac{N_R^{\rm sf+}-N_R^{\rm sf-}}{N_{R} }\nonumber\\
&=&(1-2w_{\rm sf}^+)\frac{N_{R}^+}{N_R}-(1-2w_{\rm sf}^-)\frac{N_{R}^-}{N_R} \nonumber\\
&=&(1-2w_{\rm sf}^+)\frac{w_0+w_s}{w_0+S_iw_s}\frac{1+S_i}{2}\nonumber\\
&-&(1-2w_{\rm sf}^-)\frac{w_0-w_s}{w_0+S_iw_s}\frac{1-S_i}{2}\nonumber\\
&=& \frac{w_s+S_iw_0}{w_0+S_iw_s}\nonumber\\
&-&2w_{\rm sf}^+\frac{w_0+w_s}{w_0+S_iw_s}\frac{1+S_i}{2}+2w_{\rm sf}^-\frac{w_0-w_s}{w_0+S_iw_s}\frac{1-S_i}{2}.\nonumber\\\label{SRf}
\end{eqnarray}
After the interaction, the average spin of the total beam, taking into account  all polarization effects, will be
\begin{eqnarray}
 S_{\rm tot} &=&\frac{S_{R}^{\rm sf} N_R+S_{NR}N_{NR}}{N}\nonumber\\
&=& S_i-2w_{\rm sf}^+(w_0+ w_s)\frac{1+S_i}{2}+2w_{\rm sf}^-(w_0- w_s)\frac{1-S_i}{2}.\nonumber\\\label{sf-si}\
\end{eqnarray}
Thus,  the change of the total beam polarization $ \Delta S_{\rm tot} \equiv  S_{\rm tot} -S_i$ is only due to spin-flip as it is $\propto w_{\rm sf}^\pm$, while the NRP is canceled by the similar contribution for the radiating electrons, see Eqs.~(\ref{DSF})-(\ref{DSNR}).
Introducing radiation probabilities with the spin-flip $W_{R}^{\rm sf\pm} \equiv  w_{\rm sf}^\pm w_R^\pm$, Eq.~(\ref{sf-si}) reads:
\begin{eqnarray}
  \Delta S_{\rm tot} &=&-2W_{R}^{\rm sf+}\frac{1+S_i}{2}+2W_{R}^{\rm sf-}\frac{1-S_i}{2}. \label{stot} 
\end{eqnarray}

The expressions for $w_{\rm sf}^\pm$ can be found from the fully spin resolved emission probability \cite{Liyf2020}: 
\begin{eqnarray}\label{WR1}
 W_R   &=&\Delta u \Delta t \frac{C_R}{2}(a_{1}+{\bf a}_{2}\cdot{\bf S}_{i}+{\bf b}\cdot{\bf S}_f),
\end{eqnarray}
 with \begin{eqnarray}\label{F00}
\label{b}
{\bf b}&=& b_0\mathbf{S}_i+\mathbf{b}_1+b_2({\bf S}_i\cdot\hat{{\bf e}}_v)\hat{{\bf e}}_v,\label{bbb}\\
b_0 &=& -(1+u)\left[{\rm IntK}_{\frac{1}{3}}(u')-2{\rm K}_{\frac{2}{3}}(u')\right]\\
\mathbf{b}_1&=&-u(1+u){\rm K}_{\frac{1}{3}}(u')\hat{{\bf e}}_2\\
b_2&=&-u^2\left[{\rm IntK}_{\frac{1}{3}}(u')-{\rm K}_{\frac{2}{3}}(u')\right],
\end{eqnarray}
and ${\bf S}_f$ being the final spin.  Please refer to Eq.(\ref{W1}) for  $a_{1}$ and ${\bf a}_{2}$.
Comparing the two expressions for the photon emission probability with spin-flip 
\begin{eqnarray}
W_{R}^{\rm sf}\equiv W_R(\mathbf{S}_f)|_{\mathbf{S}_f=-\mathbf{S}_i}=w_{\rm sf}(\mathbf{S}_i)w_R,
\end{eqnarray}
which reads $a_1+\mathbf{a}_2\cdot \mathbf{S}_i-\mathbf{b}\cdot \mathbf{S}_i=2(a_1+\mathbf{a}_2\cdot \mathbf{S}_i)w_{\rm sf}(\mathbf{S}_i)$,
we obtain for the spin flip conditional probability:
\begin{eqnarray}
 w_{\rm sf}(\mathbf{S}_i)=\frac{1}{2}\left( 1-\frac{\mathbf{b}\cdot \mathbf{S}_i}{a_1+\mathbf{a}_2\cdot \mathbf{S}_i}\right). 
\end{eqnarray}
When using the quantization axis along $-\hat{{\bf e}}_2$ we obtain
\begin{eqnarray} w_{\rm sf}^\pm  =\frac{1}{2}\left( 1-\frac{b_0+b'_2\pm b_1}{a_1\pm a_2 }\right),\label{wsfpm}
\end{eqnarray}
with $b_2'=b_2(\hat{{\bf e}}_2 \cdot \hat{{\bf e}}_v)^2$.
From Eq.~(\ref{stot}) and using Eq.~(\ref{wsfpm}) the change of the total beam polarization due to radiation is:
\begin{eqnarray}
\Delta \mathbf{S}_{\rm tot} &=& \Delta u \Delta t C_R(\mathbf{b}-\mathbf{S}_ia_1-\mathbf{a}_2).\label{STOT}
\end{eqnarray}
From Eqs.(\ref{SRf}) and (\ref{wsfpm}) [or alternatively Eq.~(\ref{WR1})], the average polarization of the radiating  electrons due to all effects is derived:
\begin{eqnarray}
 \Delta \mathbf{S}_R&=& \left(\frac{\mathbf{b}}{a_1+\mathbf{a}_2\cdot \mathbf{S}_i}-\mathbf{S}_i\right) \frac{N_R}{N}\nonumber\\\label{DSR}
&=&\Delta u \Delta t C_R\left[\mathbf{b}-\mathbf{S}_i(a_1+\mathbf{a}_2\cdot \mathbf{S}_i)\right].
\end{eqnarray}
We confirm via Eqs.~(\ref{DSNR}) and (\ref{DSR}) that $\Delta \mathbf{S}_{\rm tot}=\Delta \mathbf{S}_{R}+ \Delta \mathbf{S}_{NR}$. The spin-flip induced part of the radiative polarization  can be singled out using Eq.~(\ref{DSF}):
 \begin{eqnarray}
 \Delta \mathbf{S}_R^{\rm sf}&=&  \Delta \mathbf{S}_R  - \Delta \mathbf{S}_R^{\rm no-sf}\nonumber\\
&=&\Delta u \Delta t C_R(\mathbf{b}-\mathbf{S}_ia_1-\mathbf{a}_2) .
\end{eqnarray}
We stress that  $ \Delta \mathbf{S}_R^{\rm sf}$  coincides with the total beam polarization change of Eq.~(\ref{STOT}), which indicates again that the total beam polarization arises solely due to the spin-flip.
The total spin  change of Eq.~(\ref{STOT}) is in accordance with the total quantum probability given in Sec.~\ref{QED}. 

\section{Radiative polarization within nonlinear QED theory}\label{QED}

\subsection{The QED treatment of radiative polarization}

Here we collect the information on the spin resolved photon emission probabilities derived with the Baier-Katkov QED operator  method in LCFA \cite{Baier1998}. 

In LCFA, the photon emission probability is determined by the local value of the quantum parameter $\chi_{e}$. LCFA  is generally considered as a good approximation if the formation length for radiation is
 much smaller than the laser wavelength and the typical size of the electron  trajectory
 \cite{Ritus1985,Baier1998}. This is usually the case for ultraintense laser fields with  $a_0\gg 1$. Here, $a_0 \equiv |e|E_0/(m\omega_0) $ is the invariant laser field parameter with  $E_0$ denoting the laser field amplitude and $\omega_0$ the laser frequency.   

The fully spin resolved photon emission probability, summed up by the emitted photon polarization,  is given by Eq.~(\ref{WR1}).
 After the photon emission, the electron spin is in a mixed state with  
\begin{eqnarray}\label{RP}
{\bf S}_R&=&\frac{\bf b}{a_1+{\bf a}_{2}\cdot{\bf S}_{i}},
\end{eqnarray}
determined from Eq.(\ref{WR1}).

\subsection{The QED treatment of non-radiative polarization}\label{OLSE_calc}

The non-radiative polarization in QED up to the first order  in $\alpha$ is given by the sum of the propagation diagram and the interference one of the OLSE with the propagation amplitude, see Fig.~1. We present the QED calculations of Ref.~\cite{Torgrimsson_2021}, and check its agreement with NRP via the simple man's model. 
The $\alpha^0$-order spin-dependent probability comes from propagation diagram and reads
\begin{equation}
W^{(0)}=\frac{1}{2}\left(1+\mathbf{S}_{f}\cdot \mathbf{S}_{i}\right).
\end{equation}
The $\alpha$-order loop contribution to spin variation comes from the product of the tree-level propagation diagram and one-loop propagation diagram. The first-order loop probability takes the form:
\begin{equation}
W^{(L)}=2\mathbf{\textrm{Re}}M_{0}M_{1}^*,
\end{equation}
where $M_{0}=\frac{1}{2}\bar{u}_{f}u_{i}$ is the zeroth order scattering amplitude, and
$M_{1}$  the first order  scattering amplitude (Eq. (B11) in \cite{Torgrimsson_2021}).  Within the LCFA, the probability reads
\begin{align}\label{OLSE_Greger}
W^{(L)} &=\left\langle W^{(L)}\right\rangle +W_{0}^{(L)}\cdot S_{i}+W_{1}^{(L)}\cdot S_{f}+S_{f}\cdot W_{10}^{(L)}\cdot S_{i},
\end{align}
where
\begin{align*}
&\left\{ \left\langle W^{(L)}\right\rangle ,W_{0}^{(L)},W_{1}^{(L)},W_{10}^{(L)}\right\} \\  &=\frac{\alpha}{2}\int\frac{d\sigma}{kp}\int_{0}^{1}ds\left\{ \left\langle R^{(L)}\right\rangle ,R_{0}^{(L)},R_{1}^{(L)},R_{10}^{(L)}\right\}
\end{align*}
with $s=\frac{1}{1+u},ds=-\frac{1}{\left(1+u\right)^{2}}du,\sigma=\left(\phi_{1}+\phi_{2}\right)/2,dt=\frac{\gamma }{kp}d\sigma$, and
\begin{align}\nonumber\label{RL}
\left\langle R^{(L)}\right\rangle  & =\textrm{A\ensuremath{\textrm{i}_{1}}}\left(\xi\right)+\kappa\frac{\textrm{Ai}'\left(\xi\right)}{\xi},\\\nonumber
R_{0}^{(L)} & =R_{1}^{(L)}=-q\frac{\textrm{Ai}\left(\xi\right)}{\sqrt{\xi}}\boldsymbol{\hat{B}},\\
R_{10}^{(L)} & =\left\langle R^{(L)}\right\rangle \boldsymbol{I}+q\frac{\textrm{Gi}\left(\xi\right)}{\sqrt{\xi}}\left(\boldsymbol{\hat{E}}\boldsymbol{\hat{k}}-\boldsymbol{\hat{k}}\boldsymbol{\hat{E}}\right).
\end{align}
Here $\kappa=1/s+s=\frac{2+2u+u^{2}}{1+u},r=1/s-1=u,q=\frac{u}{1+u},\xi=\left(\frac{r}{\chi}\right)^{2/3}=\left(\frac{u}{\chi}\right)^{2/3}$.

To prove that the QED calculation based on the loop diagrams Eq.~(\ref{OLSE_Greger}) is equivalent to the NRP probability of Eq.~(\ref{NR}), we express the Airy functions via modified Bessel functions with the argument of $u'=\frac{2}{3}\frac{u}{\chi}$:
\begin{align*}
\textrm{Ai}\left(\xi\right) & =\frac{1}{\pi\sqrt{3}}\sqrt{\xi}\textrm{K}_{1/3}\left(u'\right),\\
\textrm{A\textrm{i}}_{1}\left(\xi\right) & =\frac{1}{\pi\sqrt{3}}\textrm{IntK}_{\frac{1}{3}}\left(u'\right),\\
\textrm{Ai}'\left(\xi\right) & =-\frac{\xi}{\pi\sqrt{3}}\textrm{K}_{\frac{2}{3}}\left(u'\right).
\end{align*}
Thus, $\left\langle R^{(L)}\right\rangle, R_{0}^{(L)},R_{1}^{(L)},R_{10}^{(L)}$ in Eq. (\ref{RL}) can be written as
\begin{widetext}
\begin{align}\nonumber\label{PRL}
\left\langle R^{(L)}\right\rangle  & =-\frac{1}{\pi\sqrt{3}}\frac{1}{1+u}\left[-\left(1+u\right)\textrm{IntK}_{\frac{1}{3}}\left(u'\right)+\left(2+2u+u^{2}\right)\textrm{K}_{\frac{2}{3}}\left(u'\right)\right]\\\nonumber
R_{0}^{(L)} & =R_{1}^{(L)}=\frac{1}{1+u}\frac{1}{\pi\sqrt{3}}u\textrm{K}_{\frac{1}{3}}\left(u'\right)\mathbf{\hat{e}}_{2}\\\nonumber
R_{10}^{(L)} & =-\frac{1}{\pi\sqrt{3}}\frac{1}{1+u}\left[-\left(1+u\right)\textrm{IntK}_{\frac{1}{3}}\left(u'\right)+\left(2+2u+u^{2}\right)\textrm{K}_{\frac{2}{3}}\left(u'\right)\right]\boldsymbol{I}\\
&+\frac{u}{1+u}\frac{\textrm{Gi}\left(\xi\right)}{\sqrt{\xi}}\left(\mathbf{\hat{e}}_{1}\mathbf{\hat{e}}_{v}-\mathbf{\hat{e}}_{v}\mathbf{\hat{e}}_{1}\right).
\end{align}
\end{widetext}
Note that $\mathbf{\hat{e}}_{1}\mathbf{\hat{e}}_{v}-\mathbf{\hat{e}}_{v}\mathbf{\hat{e}}_{1}$
is an off-diagonal matrix that leads to spin rotation in $\mathbf{\hat{e}}_{1},\mathbf{\hat{e}}_{v}$ plane and the integration of the coefficient over $u$ gives $\int\frac{u}{\left(1+u\right)^{3}}du\frac{\textrm{Gi}\left(\xi\right)}{\sqrt{\xi}}=\mu\frac{\chi}{\alpha}$
with $\mu=\frac{g-2}{2}$ being the anomalous magnetic moment (Eq. (25) in \cite{Ilderton2020}). Substituting Eq. ({\ref{PRL}}) into Eq.(\ref{OLSE_Greger}), we obtain the first-order probability coming from interference diagram
\begin{widetext}
\begin{align}
W^{(L)}=&-\frac{1}{2}\Delta t\int C_{R}du\left[-\left(1+u\right)\textrm{IntK}_{\frac{1}{3}}\left(u'\right)+\left(2+2u+u^{2}\right)\textrm{K}_{\frac{2}{3}}\left(u'\right)-u\textrm{K}_{\frac{1}{3}}\left(u'\right)\mathbf{\hat{e}}_{2}\cdot\boldsymbol{S}_{i}\right]\nonumber\\&+\frac{1}{2}\Delta t\int C_{R}du\left\{u\textrm{K}_{\frac{1}{3}}\left(u'\right)\mathbf{\hat{e}}_{2}-\left[-\left(1+u\right)\textrm{IntK}_{\frac{1}{3}}\left(u'\right)+\left(2+2u+u^{2}\right)\textrm{K}_{\frac{2}{3}}\left(u'\right)\right]\boldsymbol{S}_{i}\right\}\cdot\boldsymbol{S}_{f}\nonumber\\
&+\frac{\mu}{2}\omega_{0}a_{0}\left(\varphi\right)\Delta t\left[\left(\boldsymbol{S}_{f}\cdot\mathbf{\hat{e}}_{1}\right)\left(\boldsymbol{S}_{i}\cdot\mathbf{\hat{e}}_{v}\right)-\left(\boldsymbol{S}_{f}\cdot\mathbf{\hat{e}}_{v}\right)\left(\boldsymbol{S}_{i}\cdot\mathbf{\hat{e}}_{1}\right)\right]. \label{PL1}\\
W&=W^{(0)}+W^{(L)}=  W_{NR} +\frac{\mu}{2}\omega_{0}a_{0}\left(\varphi\right)\Delta t\left[\left(\boldsymbol{S}_{f}\cdot\mathbf{\hat{e}}_{1}\right)\left(\boldsymbol{S}_{i}\cdot\mathbf{\hat{e}}_{v}\right)-\left(\boldsymbol{S}_{f}\cdot\mathbf{\hat{e}}_{v}\right)\left(\boldsymbol{S}_{i}\cdot\mathbf{\hat{e}}_{1}\right)\right], \label{PL2}
\end{align}
\end{widetext}
 where the probability for no photon emissions reads:
\begin{equation}
\label{NR}
W_{NR}=\frac{1}{2}\left(c+{\bf S}_{f} \cdot{\bf d}\right),
\end{equation}
where,
$c=1-C_R(a_{1}+{\bf a}_{2}\cdot{\bf S}_{i})$ and ${\bf d}={\bf S}_{i}\left(1-\int C_Ra_1 du \Delta t\right)-\int C_R{\bf a}_2 du \Delta t$
The average spin due to NRP is:
\begin{eqnarray}\label{NP}
{\bf S}_{NR}&=&{\bf d}/c.
\end{eqnarray}
The expression for NRP probability $W_{NR}$ via OLSE interference diagram $W = W^{(0)} + W^{(L)}$ of Eq.~(\ref{PL2}) is exactly the same as that deduced from the simple man's model, see $W_{NR}$ of Eq.~(\ref{NR3}), except for an extra rotating term [the last term proprtional to $\mu$ in Eq.~(\ref{PL2})], describing the spin precession modification due to the anomalous magnetic moment. The latter we account for in our Monte Carlo code using the anomalous magnetic moment in the BMT equation, see Appendix A.


The total fully spin resolved probability of the interaction is derived via combining  of radiative and non-radiative probabilities of Eqs.~(\ref{WR1}) and (\ref{NR})
\begin{eqnarray}\label{wtot}
dW_{\rm tot} &=& dW_R+W_{NR}\\\nonumber
 &=& \frac{1}{2}\left[1+\mathbf{S}_f\cdot\left(\mathbf{b}+\mathbf{S}_i-a_1\mathbf{S}_i+\mathbf{a}_2\right)C_R\Delta u\Delta t \right].
\end{eqnarray}
 The latter yield the same final average spin which has been deduced from the simple man's model, see Eq.~(\ref{STOT}) in Sec.~\ref{simpleman}.

\subsection{The radiative spin dynamics  }

 In this section we derive differential equations describing the evolution of the polarization dynamics. 
 
\begin{eqnarray}\label{Svar}
\Delta{\bf S}_{\rm tot} & =&\sum_{S_f}\left(\int\frac{dW_R}{dudt}{\bf S}_{R}du\Delta t-{\bf S}_{i}+W_{NR}{\bf S}_{NR}\right),\\
 & =&\sum_{S_f}\left[\int\frac{dW_R}{dudt}\left({\bf S}_{R}-{\bf S}_{i}\right)du\Delta t+W_{NR}\left({\bf S}_{NR}-{\bf S}_{i}\right)\right],\nonumber
\end{eqnarray}
where we have used $W_{NR}{\bf S}_{i}+\int\sum_{S_f}\frac{dW_R}{dudt}{\bf S}_{i}du\Delta t={\bf S}_{i}$.
One can split the spin variation in Eq. (\ref{Svar}) into the radiative polarization term and the nonradiative (self-energy) term. The radiative polarization term takes the form:
\begin{eqnarray}
\Delta{\bf S}_R & =&\sum_{S_f}\left[\int\frac{dW_R}{dudt}\left({\bf S}_{R}-{\bf S}_{i}\right)du\Delta t\right],\\
 & =&\int C_{R}\left(a_{1}+{\bf a}_{2}\cdot{\bf S}_{i}\right)\left(\frac{{\bf b}}{a_{1}+{\bf a}_{2}\cdot{\bf S}_{i}}
 -{\bf S}_{i}\right)du\Delta t.\nonumber
\end{eqnarray}
Then,
\begin{widetext}
\begin{eqnarray}\label{dSRdt}
\frac{d{\bf S}_R}{dt} & =&\int C_{R}\left[{\bf b}-{\bf S}_{i}\left(a_{1}+{\bf a}_{2}\cdot{\bf S}_{i}\right)\right]du\nonumber\\
 &=&\int C_{R}\left\{ \left[-(1+u)[{\rm IntK}_{\frac{1}{3}}(u')-2{\rm K}_{\frac{2}{3}}(u')]{\bf S}_{i}-u(1+u){\rm K}_{\frac{1}{3}}(u')\hat{{\bf e}}_{2}\right]\right.\nonumber\\
&& \left.-{\bf S}_{i}\left[-(1+u){\rm IntK}_{\frac{1}{3}}(u')+(u^{2}+2u+2){\rm K}_{\frac{2}{3}}(u')-{\bf S}_{i}\cdot\hat{{\bf e}}_{2}u{\rm K}_{\frac{1}{3}}(u')\right] \right\} du\nonumber\\
 & =&-\int C_{R}\left\{ u(1+u){\rm K}_{\frac{1}{3}}(u')\hat{{\bf e}}_{2}+{\bf S}_{i}\left[u^{2}{\rm K}_{\frac{2}{3}}(u')-\left({\bf S}_{i}\cdot\hat{{\bf e}}_{2}\right)u{\rm K}_{\frac{1}{3}}(u')\right]\right\} du.
\end{eqnarray}
\end{widetext}
The nonradiative spin variation induced by self-energy is
\begin{eqnarray}
\Delta{\bf S}_{NR} & =&W_{NR}\left({\bf S}_{NR}-{\bf S}_{i}\right)\nonumber\\
 & =&c\times\left(\frac{\bf d}{c}-{\bf S}_{i}\right)\nonumber\\
 & =&\left[{\bf S}_{i}\int C_{R}\left({\bf a}_{2}\cdot{\bf S}_{i}\right)du-\int C_{R}{\bf a}_{2}du\right]\Delta t,
\end{eqnarray}
such that
\begin{eqnarray}\label{dSNR}
\frac{d{\bf S}_{NR}}{dt} & =&{\bf S}_{i}\int C_{R}\left({\bf a}_{2}\cdot{\bf S}_{i}\right)du-\int C_{R}{\bf a}_{2}du\nonumber\\
 & =&-{\bf S}_{i}\int C_{R}\left({\bf S}_{i}\cdot\hat{{\bf e}}_{2}\right)u{\rm K}_{\frac{1}{3}}(u')du+\int C_{R}\hat{{\bf e}}_{2}u{\rm K}_{\frac{1}{3}}(u')du.\nonumber\\
\end{eqnarray}
 The Baier's equation for the average spin evolution in the limit of $\chi\ll 1$, see Eq.~(3.23) in Ref.~\cite{Baier1972}, is recovered via 
\begin{eqnarray}
\frac{d{\mathbf S}_{\rm tot}}{dt}=\frac{d{\bf S}_{R}}{dt}+\frac{d{\bf S}_{NR}}{dt},
\end{eqnarray}
using corresponding asymptotic expressions in Eqs.~(\ref{dSRdt}) and (\ref{dSNR}).

In Ref. \cite{Baier_b_1973} Baier et al. derive an analogous equation for the average electron momentum $\bm{p}$, which takes into account the emission of photons at the leading order in $\alpha$ and which in our notation reads
\begin{equation}
\label{dp_Baier}
\frac{d\bm{p}}{dt}=-\int dW_R(\bm{k}) \bm{k}+\left.\frac{d\bm{p}}{dt}\right|_L,
\end{equation}
where $dW_R(\bm{k})$ is the probability per unit time of the emission of a photon with momentum between $\bm{k}$ and $\bm{k}+d\bm{k}$, and $d\bm{p}_L/dt|_L$ is the variation of the momentum in the absence of radiation, i.e., the Lorentz force. The first term in Eq. (\ref{dp_Baier}) corresponds to the radiative change $d\bm{S}_R/dt$ in the case of the spin, whereas the second one corresponds to the BMT equation for the spin. This shows that, unlike the average spin, the average momentum of the electron changes either because of the Lorentz force or because of the radiation of photons.

 We employ a QED Monte Carlo simulation code similar to Ref.~\cite{li2022helicity}, which includes the three sources of strong field polarization, see~Appendix A:  the RP is accounted for by the spin-resolved QED probability in the local constant field approximation \cite{Baier1998,chen2022}, the  NRP by the corresponding probability \cite{CAIN,Liyfei2020}, and the OLSE contribution to the anomalous magnetic moment by the appropriate modification of BMT equation.

\section{The OLSE signature in electron angular distribution}\label{result}

\subsection{M{\o}ller polarimetry}\label{Moller}

The most relevant technique for the detection of  $\sim 100 $ MeV electron polarization is M{\o}ller polarimetry.
The electron polarization is defined by the vector $\mathbf{S}_i$, which is the electron spin in the rest frame of the electron. For a monoenergetic electron beam, the average beam polarization is straightforwardly the average over $\mathbf{S}_i$:
\begin{eqnarray}
\langle \mathbf{S}\rangle=\frac{\sum_i n_i \mathbf{S}_i}{\sum_i n_i},
\label{zeta}
\end{eqnarray}
with the number of the electrons $n_i$ in the beam with polarization $\mathbf{S}_i$, because the rest frame of all electrons is the same. In this case polarimetry determines the average polarization $\langle \mathbf{S}\rangle $ defined by Eq.~(\ref{zeta}).

However, in our setup, the electrons after the interaction are distributed in a quite large energy and angle range. 
In this case, the rest frames of electrons are different and the definition of Eq.~(\ref{zeta})  is not relevant. How can in this case M{\o}ller polarimetry be applied, and which parameter will be measured?  Let us calculate the signal of M{\o}ller polarimetry with our broad electron distribution. 
We will assume that the electrons  collected at certain angle region are focused via beam optics over the angle, but still have a large energy distribution  $\Delta \varepsilon \sim \varepsilon_0$ around the mean energy $\varepsilon_0$.

M{\o}ller polarimetry employs the scattering of polarized solid-targets off the electrons. The difference of the scattered electron yield ${\cal N}_e^+-{\cal N}_e^-$ is measured between the cases when the electron helicity is parallel or anti-parallel to the target polarization direction. It allows one to derive the average polarization vector from the incoming particles' polarization resolved  M{\o}ller scattering cross-section.  The cross-section in the center of momentum frame reads \cite{Moller1932,Arrington1992}:
\begin{eqnarray}
\frac{d\sigma}{d\Omega'}= \frac{d\sigma_0}{d\Omega'}\left( 1+\sum_{i,j} P_B^i A_{i,j} P_T^j \right),
\label{sigma}
\end{eqnarray}
where $P_B^i (P_T^j)$ are the components of the beam (target) polarization, as measured in the rest frame of the
beam (target) electrons. Here, we set a new coordinate system with $z'$-axis along the momentum of the electron beam to be detected and the $y'$-axis normal to the M{\o}ller scattering plane, and add a superscript of $'$ into the symbols of angles to distinguish with those in the $xyz$-coordinate system we used for the laser-electron interaction. 

The cross section is characterized by the unpolarized cross section $\frac{d\sigma_0}{d\Omega'}$, and nine
asymmetries $A_{i,j}$. By measuring the spin-dependent cross section on a target of known polarization $P_T$,
Eq. ~(\ref{sigma}) can be used to extract the beam polarization components $P_B^i$. To lowest order in QED and
using the ultra-relativistic approximations, the  unpolarized cross section and nine asymmetries are  \cite{Moller1932,Arrington1992}:
\begin{eqnarray}
\frac{d\sigma_{0}}{d\Omega'}&=&\left[\frac{\alpha\left(1+\cos\theta'_{\text{CM}}\right)\left(3+\cos^{2}\theta'_{\text{CM}}\right)}{2m\sin^{2}\theta'_{\text{CM}}}\right]^{2},\\
A_{z'z'}&=-&\frac{(7+{\rm cos}^2 \theta'_{\rm CM}) {\rm sin}^2 \theta'_{CM}}{(3+{\rm cos}^2 \theta'_{\rm CM})^2},\\
-A_{x'x'}&=&A_{y'y'}=\frac{{\rm sin}^4 \theta'_{\rm CM} }{(3+{\rm cos}^2 \theta'_{\rm CM})^2},\\
A_{x'z'}&=&A_{z'x'}=-\frac{2{\rm sin}^3 \theta'_{\rm CM}{\rm cos} \theta'_{\rm CM} }{\gamma (3+{\rm cos}^2 \theta'_{\rm CM})^2},\\
A_{x'y'}&=&A_{y'x'}=A_{y'z'}=A_{z'y'}=0.
\end{eqnarray}
Note that $\theta'_{\rm CM}$ is the center of mass (CM) scattering angle. To measure the longitudinal polarization,  the experimentally determined quantity is the asymmetry parameter of
\begin{eqnarray}
{ \cal A}&=&\frac{{\cal N}_e^+-{\cal N}_e^-}{{\cal N}_e^++{\cal N}_e^-}.
\label{asy}
\end{eqnarray}
This asymmetry parameter is related to the theoretical asymmetry by
\begin{eqnarray}
{ \cal A}&=& P_B P_T A_{z'z'}.
\label{asy1}
\end{eqnarray}

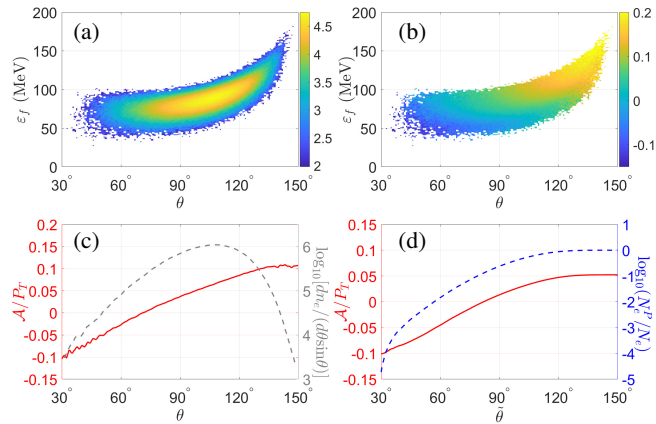
\begin{figure}[b]
 	\includegraphics[width=1.0\linewidth]{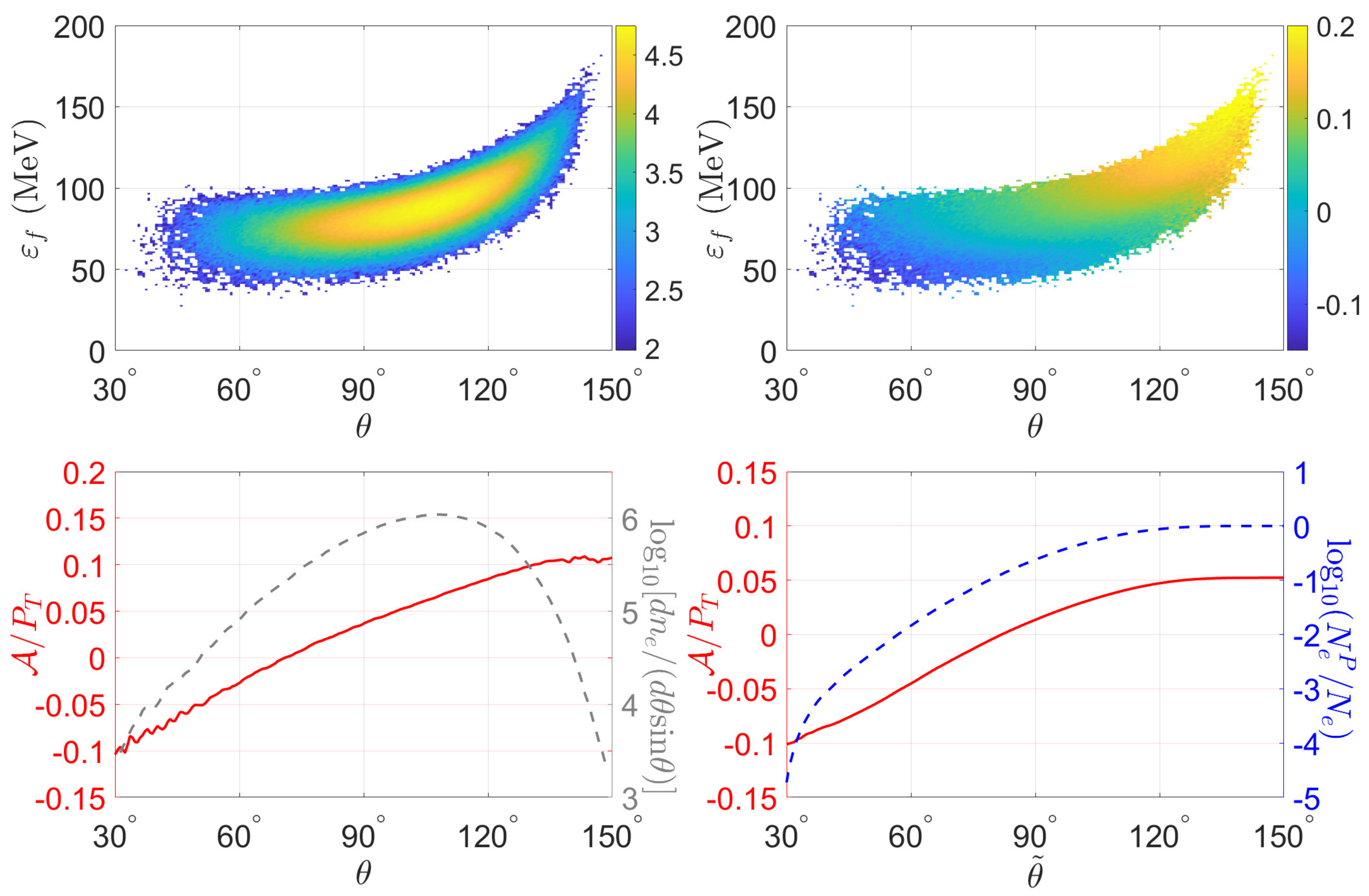}
  \begin{picture}(300,-30)
  \put(25,157){\small (a)}
  \put(148,157){\small(b)}
  \put(25,78){\small (c)}
  \put(148,78){\small(d)}
 	  \end{picture}
		  \caption{(a) Electron number density log$_{10}$[d$n_{e}$/($\sin\theta$d$\theta$d$\varepsilon_f$)], and (b) Longitudinal polarization $P_\parallel$ vs $\varepsilon_f$ and polar angle of $\theta={\rm acos} (p_z/\gamma)$;  (c) The asymmetry parameter ${\cal A}$ of M\o{}ller polarimetry averaged over the energy (with the target polarization $P_T$) and number density   log$_{10}$[$d n_{e}$/($d\theta {\rm sin}\theta)$] vs $\theta$; (d) ${\cal A}$ of the selected electrons with $\theta<\tilde{\theta}$ (left axis) and their relative fraction (right axis) vs $\tilde{\theta}$ at a detection angle of $\theta'_d=2/15$.} \label{fig2}
\end{figure}

    \begin{figure}      	
\includegraphics[width=0.8\linewidth]{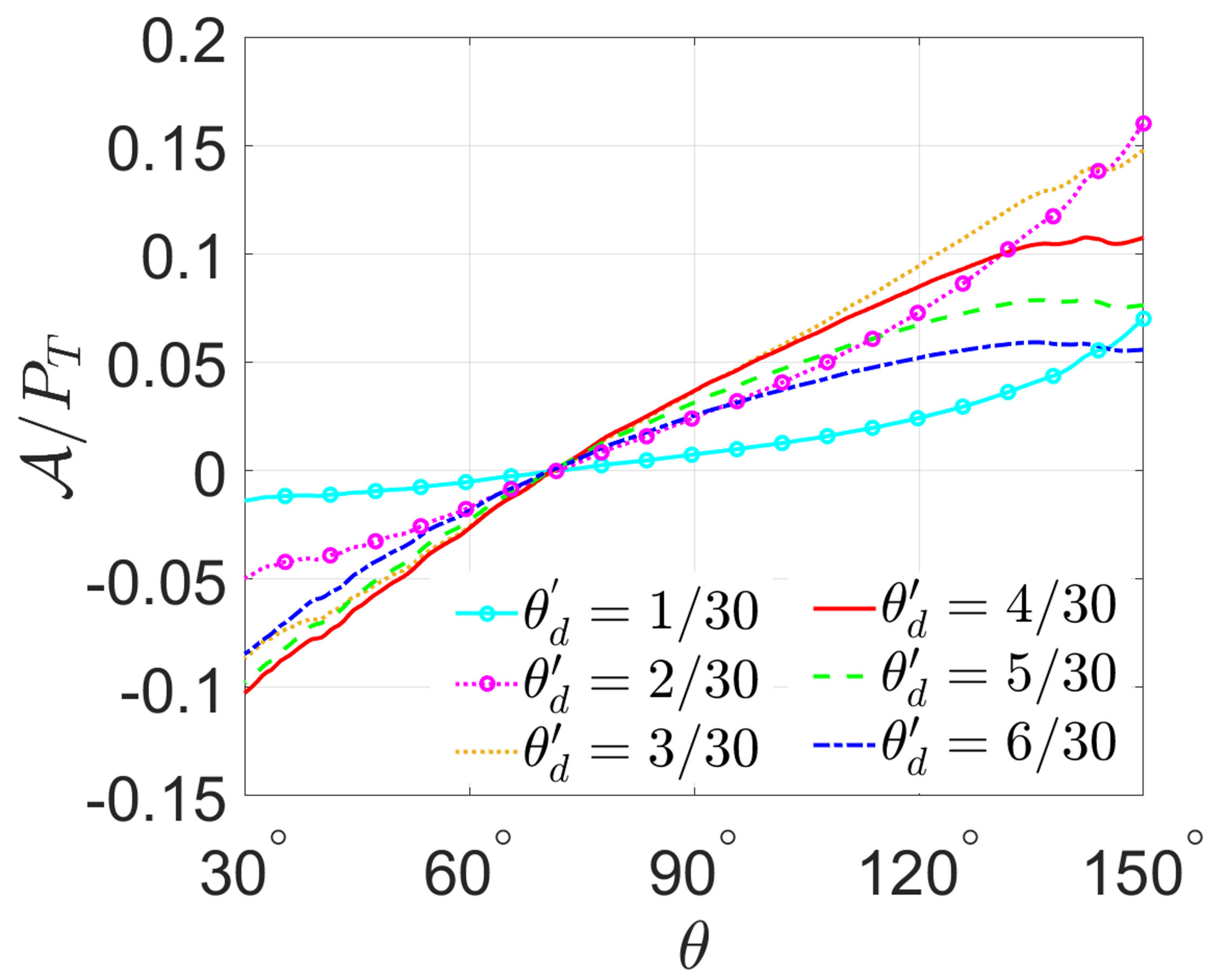}  
\begin{picture}(300,0) 
\end{picture}	  
\caption{ The average characteristic spin-parameter ${\cal A}/P_T $ vs $\theta$ for detection angle $\theta'_d$ changing from 1/30 to 6/30.}  \label{moller}  
\end{figure}

In the case of a monoenergetic electron beam, with the given $P_T $ and $A_{z'z'}$, the   polarization of the beam of $ P_B$ could be deduced from the measured ${\cal A}$ via Eq.~(\ref{asy1}). As $A_{z'z'}$ is a function of $\theta'_{\rm CM}$, with the maximum value of 7/9 at $\theta'_{\rm CM}=90^\circ$, the detectors are typically located at the Lab angle of $\theta'_d$ corresponding to $\theta'_{\rm CM}=90^\circ$. The current experimental capability of measuring the asymmetry parameter is ${\cal A}_m=0.5\% \times P_T \times \frac{7}{9}=3.89\times 10^{-3} P_T$, with a given target polarization $P_T$. It leads to the experimental detection accuracy of 0.5\% in electron beam polarization.

Considering the connection between the Lab scattering angle and the center of mass scattering angle \cite{Jones2022},
 \begin{eqnarray}
 {\theta_{\rm L}^{'2} } = 2m \left(\frac{1}{p_s}- \frac{1}{p_i}\right),\\
 p_s=\frac{p_i}{2}(1+{\rm cos} \theta'_{\rm CM}),
\end{eqnarray}
the $A_{z'z'}$ is a function of the incident electron energy $m \gamma$ and  the detection angle $\theta'_d$ in the Lab frame. Here, $p_s$ ($p_i$) is the momentum of the scattered (incident) electron for M{\o}ller scattering. 

In the case of a broad energy distribution the asymmetry parameter for a certain detection angle $\theta'_d$  is given by:
 \begin{eqnarray}
 {\cal A} = P_T \frac{\sum_i \frac{d\sigma_{0}^{i}}{d\Omega_{i}'} S_{z'i} A_{z'z'}(\theta'_d, \gamma_i) n_i }{\sum_i  \frac{d\sigma_{0}^{i}}{d\Omega_{i}'} n_i } = P_T  \langle S_{z'} A_{z'z'}\rangle,
\label{zeta4}
\end{eqnarray}
where $n_i$ is the number of electrons with $\gamma_i$ and the corresponding average polarization $S_{z'i}$ for the given energy.

\subsection{Numerical QED Monte Carlo simulations}

A typical simulation result for density and polarization distributions of the scattered electrons is illustrated  in Fig.~\ref{fig2}. A left-hand CP tightly-focused Gaussian laser pulse is used. The peak laser intensity is $I_0\approx 1\times 10^{23}$ W/cm$^2$ $ (a_0=200)$, the pulse duration  $\tau=6T_0$, with period $T_0$,  wavelength  $\lambda=1 \mu$m, and focal radius $w_0=2 \lambda$. The electron bunch of a cylindrical form is considered, with a length of $L_e=3 \lambda$, and radius of $w_e=1 \lambda$,  the electron number is $N_e=7.5\times 10^5$, they are uniformly distributed longitudinally and normally distributed transversely. The initial kinetic energy is 500 MeV, the energy spread $10\%$, and the angular divergence 1 mrad.  The pair production in the applied conditions is estimated to be negligible.

The dependence of polarization and density of electrons on the azimuthal angle $\varphi$ is uniform in a circular polarized field, and only the polar angle $\theta$ for electrons with respect to the initial beam propagation direction is relevant for evidencing the role of NRP.
After the interaction, most of the electrons are reflected to the laser propagation direction ($\theta>90^\circ$), while some keep moving forward ($\theta<90^\circ$) [Fig.~\ref{fig2}(a)]. The helicity of the laser field can be transferred to the scattered electrons \cite{li2022helicity}, which is observed for the electrons with $\theta>69.7^\circ$ . The longitudinal polarization $P_\parallel$ of the forward electrons with $\theta<69.7^\circ$ is negative and opposite to that of the remaining electrons [Fig.~\ref{fig2}(b)].
 We will show below that the latter is a distinct signature of NRP.
 Through collecting electrons within a certain angle of $\theta \in [0,\tilde{\theta}]$, $\tilde{\theta}<69.7^\circ$, one could optimize the number of electrons with polarization dominated by NRP.
Our aim is to characterize the average polarization of post-selected electrons within $\theta<\tilde{\theta}$, which is a nontrivial task. The electron polarization is defined by the spin vector in the rest frame of the electron as we have mentioned. However,  in our setup, the electrons after the interaction are distributed in a large energy range, when the rest frames of electrons are different and the straightforward averaging over the spin vector is not physically valid. We characterize the polarization of the broadly distributed collection of electrons after the interaction via the polarimetry signal.

The M{\o}ller polarimetry \cite{Moller1932} is well suited to our setup with the involved energies  $\sim 100 $ MeV,  see Sec. \ref{Moller}. The M{\o}ller polarimetry measures asymmetry parameter ${\cal A}$ which is expressed by the rest frame spin vector. We calculate the average ${\cal A}$ over the energy (scaled by the target polarization $P_T$) for post-selected electrons with $\theta<\tilde{\theta}$, for each energy value using the corresponding average spin vector [Fig.~\ref{fig2}(c),(d)].
The asymmetry parameter ${\cal A}$ 
 with respect to $\theta$ for different detection angles of $\theta'_d$ is shown in Fig.\ref{moller}.  In experiment, the scaling law of ${\cal A}$ 
 over $\theta$ can be obtained by arranging a series of M{\o}ller polarimeters over angle. 
A higher ${\cal A}$ can be obtained by selecting electrons in a smaller angle region.  For instance, if we collect electrons within $[0^\circ, 40.25^\circ]$, corresponding to 0.1\% of the total, we would obtain  ${\cal A}$ about $ 0.084 P_T$, far larger than the experimental capacity of $ 3.89\times 10^{-3} P_T$ in polarization measurement \cite{Arrington1992}.

The spin dynamics during RP  process is described by the spin resolved photon emission rates of Eq. (\ref{dSRdt}), which could be approximated as:
\begin{eqnarray}\label{SFE}
\frac{d\mathbf{S}_{R}}{dt} 
&=&-\frac{\alpha m}{\sqrt{3}\pi\gamma}\int_0^\infty\frac{du}{(1+u)^3}\left\{ [u^2 \mathrm{K_{2/3}}-u\mathrm{K_{1/3}}({\mathbf S}_i \cdot \hat{\bf  e}_2 )]{\mathbf S}_i \nonumber\right.\\
&&\left.+u(1+u)\mathrm{K_{1/3}}\hat{\bf  e}_2 \right\}
\nonumber\\&\approx&-\frac{\alpha m}{\sqrt{3}\pi\gamma}\int_0^\infty\frac{du}{(1+u)^3}[u(1+u)\mathrm{K_{1/3}}\hat{\bf  e}_2],
\end{eqnarray}
The approximation is valid for initially unpolarized electrons with $|{\bf S}_{i}|\ll1$.

The polarization evolution related to NRP approximated form Eq. (\ref{dSNR}) reads:
\begin{eqnarray}\label{OLSE}\nonumber
\frac{d\mathbf{S}_{\rm NR}}{dt}&=&-\frac{\alpha m}{\sqrt{3}\pi\gamma}\int_{0}^{\infty}\frac{du}{\left(1+u\right)^{3}}\left\{ \left[\left(\mathbf{S}\cdot\hat{{\bf e}}_{2}\right)u\mathrm{K_{1/3}}\right]\mathbf{S}-u\mathrm{K_{1/3}}\hat{{\bf e}}_{2}\right\} \nonumber\\
&\approx & \frac{\alpha m}{\sqrt{3}\pi\gamma}\int_{0}^{\infty}\frac{du}{\left(1+u\right)^{3}}\left[u\mathrm{K_{1/3}}\hat{{\bf e}}_{2}\right].
\end{eqnarray}

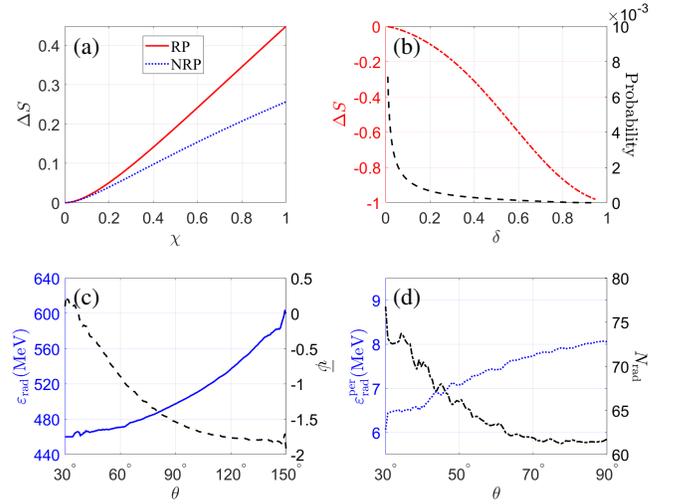
\begin{figure}
  	\includegraphics[width=1\linewidth]{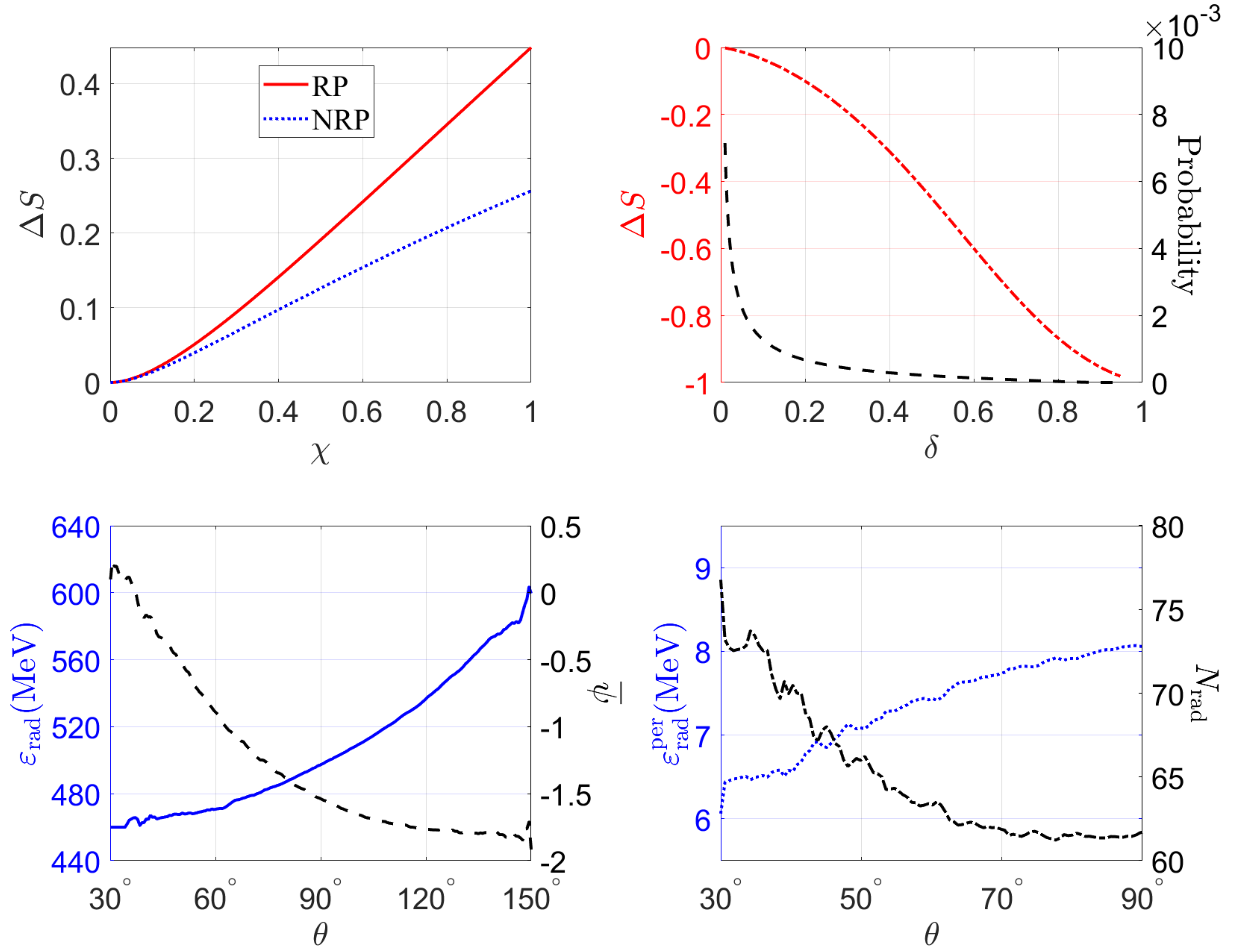}
  \begin{picture}(300,0)
  \put(25,180){\small (a)}
  \put(146,180){\small(b)}
  \put(25,85){\small (c)}
  \put(146,85){\small(d)}
 \end{picture}		
  \caption{(a) Instantaneous polarization via RP (red-solid) and NRP (blue-dotted) via Eqs.~(\ref{SFE}) and (\ref{OLSE}) vs the quantum strong-field parameter $\chi$; (b) Instantaneous polarization via RP (red-dash-dotted) and radiation probability (black-dashed) calculated from Eq.~(\ref{SFE}) vs $\delta\equiv \omega_\gamma/\varepsilon_e$  for $\chi=1$; (c) Average radiation energy loss per electron and average radiation phase $\overline{\psi}$  of multiple photon-emission events vs $\theta$. Here, $\psi$ is in unit of $2\pi$, and $\psi=\omega_0 t+k_0 z =0$ refers to the peak of the laser pulse; (d) The average radiation energy loss per photon-emission and average photon-emission times per electron vs $\theta$, for forward moving electrons. } \label{fig3}
  \end{figure}

In the colliding geometry where $\hat{\bf e}_2$ is the direction of the local magnetic field, we have $\mathbf{S}_{R}(t)\propto\mathbf{E}$ while $\mathbf{S}_{\rm NR}(t)=\propto-\mathbf{E}$, according to Eqs. (\ref{SFE}) and (\ref{OLSE}). Therefore, the RP (NRP) induces an instantaneous transverse polarization oscillating in phase with the laser field $\mathbf{E}$ ($\mathbf{-E}$) during the interaction. Since $d\mathbf{S}_{R}/dt=-d\mathbf{S}_{\rm NR}/dt\approx -\frac{\alpha m}{\sqrt{3}\pi\gamma}\int_{0}^{\infty}du\left[u\mathrm{K_{1/3}}\hat{{\bf e}}_{2}\right]$ for
 $\chi \ll 1$, the RP  cancels NRP at the leading order in $\alpha$, which causes a vanishing longitudinal polarization \cite{torgrimsson2021resummation,Torgrimsson_2021}.  However, as long as radiation reaction and  spin-flip at emission become significant at large parameter $\chi$, RP prevails over NRP [Fig.~\ref{fig3}(a)] leading to an instantaneous polarization  phase-matched with the local $\mathbf{E}$. This transverse polarization is rotated by spin precession, resulting in the helicity transfer from the laser field to the scattered electrons.
 As shown in Fig. \ref{fig3} (b), the electrons with less 
 photon recoil experience no significant spin-flip, and have smaller polarization change induced by RP. Thus, the polarization dynamics of electrons experiencing less 
 radiation recoil is dominated by NRP, which results in an opposite helicity with the laser. The radiation loss, in its turn, is correlated with the electron final propagation direction, see Fig.~\ref{fig3}(c), which contains also information on the  emission phase. Electrons undergoing significant   radiation energy loss  (also RP) at the beginning of the laser pulse tend to be reflected and accelerated to high energies by the laser field. Meanwhile, the electrons that radiate only slightly in the later part of the laser pulse, would penetrate through the laser field with a smaller $\theta$. When 
 radiative recoil is small enough, the NRP can surpass RP, resulting in electron helicity along the opposite direction, see Fig.~\ref{fig2}.

How the angle dependent 
radiation loss  is built up during multiple photon emissions is illustrated in Fig.~\ref{fig3}(d). It shows that the electrons scattered at large (small) angles have emitted less (more) photons, however, with larger (smaller) photon energy on average. This has a simple explanation.  Assuming an electron with $p_\parallel^i\approx m\gamma$, $p_\perp=0$ interacts with a CP plane-wave field.
In the laser field the electron gains a transverse momentum $\mathbf{p}_\bot =-e \mathbf{A}(\psi)$. The ultrarelativistic electron emits a photon $\omega_\gamma$ along the instantaneous momentum in the laser field, consequently, the electron final transverse momentum after the interaction will be determined by the photon recoil $p^f_{\perp}=\omega_\gamma e A(\psi_e)/\varepsilon$, with the emission phase $\psi_e$, and the electron energy in the laser field $\varepsilon=\gamma+a_0^2/4\gamma$, i.e., the deflection angle due to a single emission is $\Delta\theta\approx\sin^{-1}\,|p^f_{\perp}/\varepsilon_f|$, with the electron final energy $\varepsilon_f$, which is proportional to $\delta$. The final transverse momentum of multiple emissions is a sum of $p^f_{\perp,i}$ along random directions $\mathbf{A}(\psi_i)$, and could average out to a vanishing value at extremely large emission events.
Consequently, $\Delta \theta$ is proportional to the statistical error of the number of emissions ${N_{\rm rad}}$ during stochastic photon emissions, i.e.,  $\Delta \theta\sim 1/\sqrt{{N_{\rm rad}}}$. Thus,  statistically, $\Delta \theta$ increases with the photon energy $ \overline{\omega}_\gamma$ and decreases with $N_{rad}$, which is confirmed in Fig.~\ref{fig3}(d).\\

\section{The experimental feasibility}\label{experiment}

\subsection{Impact of laser and electron-beam parameters on the signature}
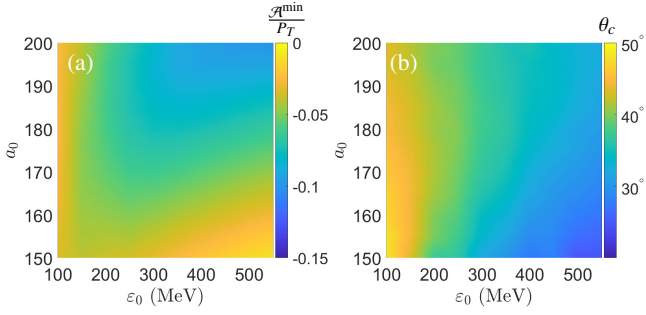
\begin{figure}[t]
  	\includegraphics[width=1\linewidth]{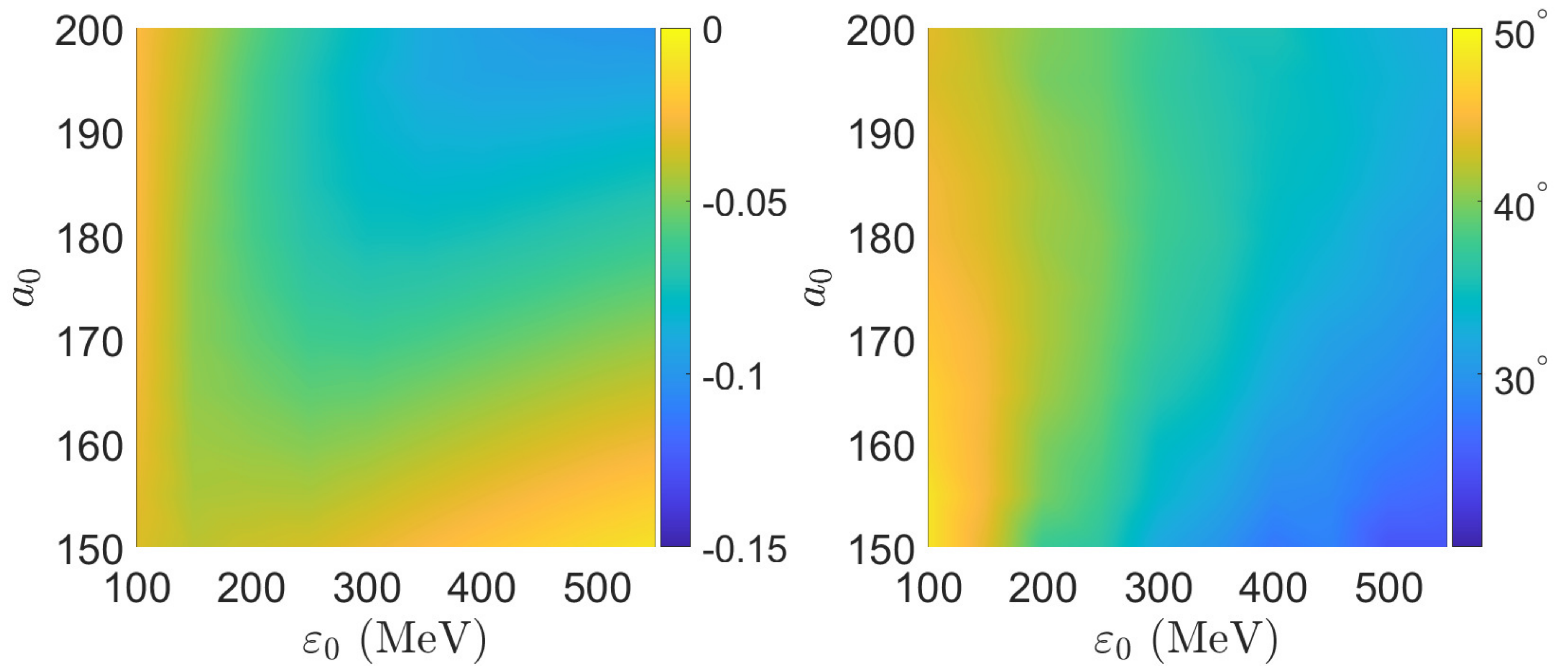}
  \begin{picture}(300,0)
  \put(25,100){\small {\color{white}(a)}}
  \put(100,117){\footnotesize $\frac{{\cal A}^{\rm min}}{P_T}$}
  \put(147,100){\small {\color{white}(b)} }
  \put(226,115){\footnotesize $\theta_c$}
 \end{picture}		
  \caption{Distributions of the minimum of ${\cal A}/P_T$ (a) and the critical angle $\theta_c$ (b) at which ${\cal A}/P_T$ changes sign, vs $a_0$ and $\varepsilon_0$. } \label{fig5}
  \end{figure}

For the experimental feasibility,  we have investigated the impact of laser intensity and initial electron  energy on the NRP signature  [Fig.~\ref{fig5}(a)].  Since the quantum parameter $\chi\sim10^{-6}\gamma a_0$ controls the strength of NRP [Fig. \ref{fig3} (a)] and the quantum stochastic effects which separate NRP and RP, the increase of $\gamma$ or $a_0$ is certainly beneficial for a stronger signature of NRP [Fig.~\ref{fig5}(a)].
Meanwhile, for a larger critical angle $\theta_c$, below which the signature of OSLE can be detected, larger  $a_0$ and smaller $\varepsilon_0$ are favorable [Fig.~\ref{fig5}(b)]. 

\begin{figure}
  	\includegraphics[width=1\linewidth]{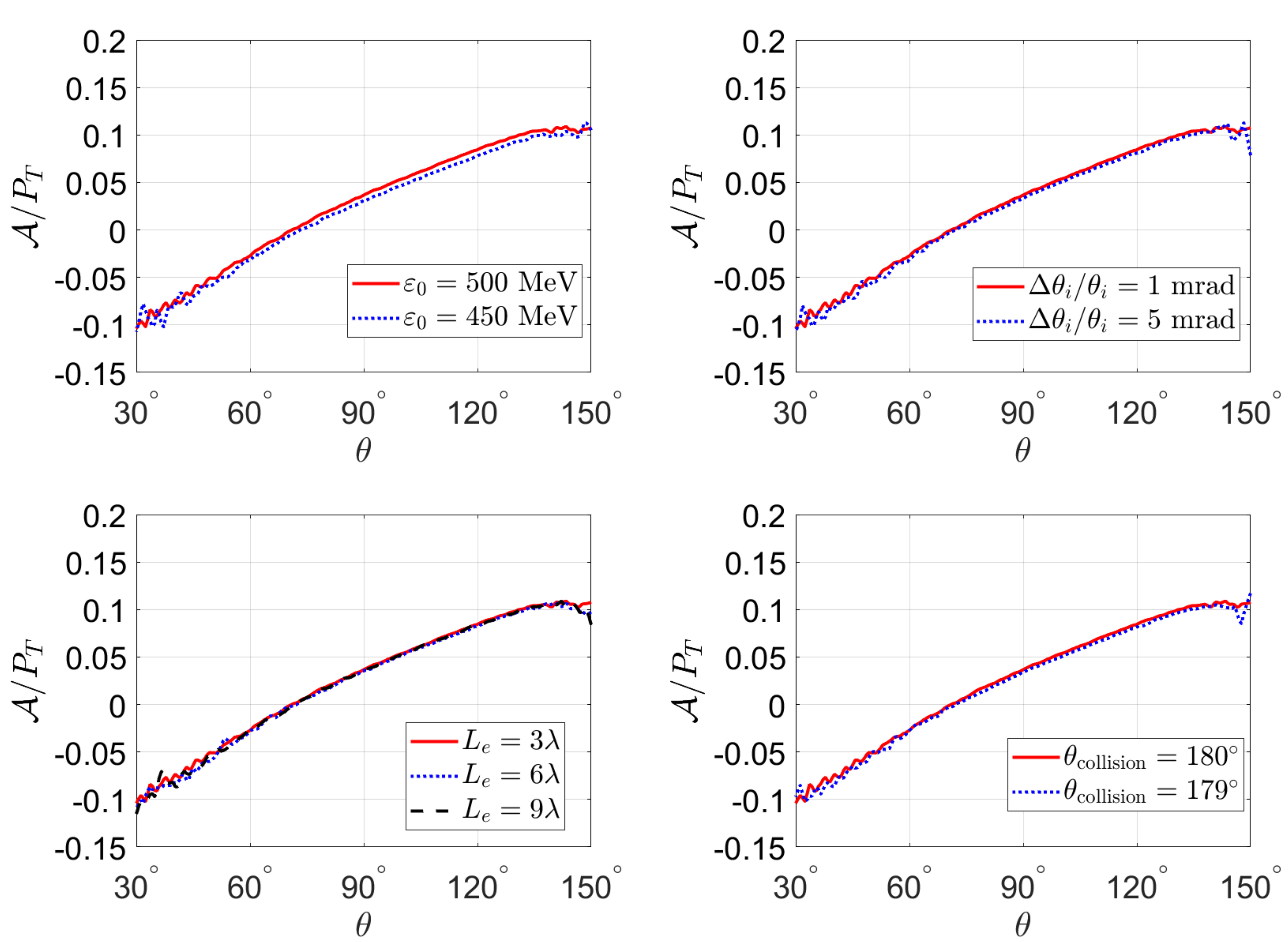}
  \begin{picture}(300,0)
  \put(28,175){\small (a)}
  \put(155,175){\small (b)}
    \put(28,84){\small (c)}
  \put(155,84){\small (d)}

 \end{picture}		
  \caption{Influence of fluctuations of electron-beam parameters, including initial electron kinetic energy $\varepsilon_0$ (a), beam divergence $\Delta \theta_i/\theta_i$ (b), bunch length $L_e$ (c) and collision angle $\theta_{\rm collision}$ (d),  on the OLSE signature. } \label{flu_1}
  \end{figure}
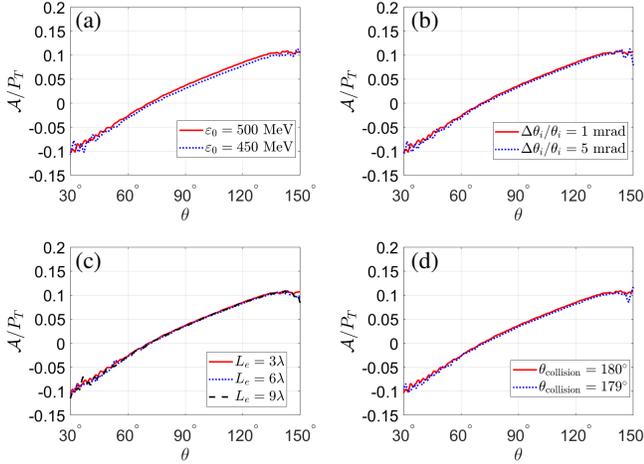

    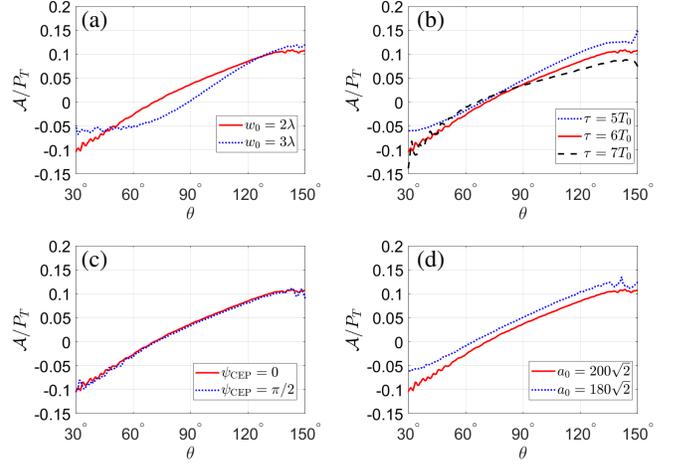
\begin{figure}
  	\includegraphics[width=1\linewidth]{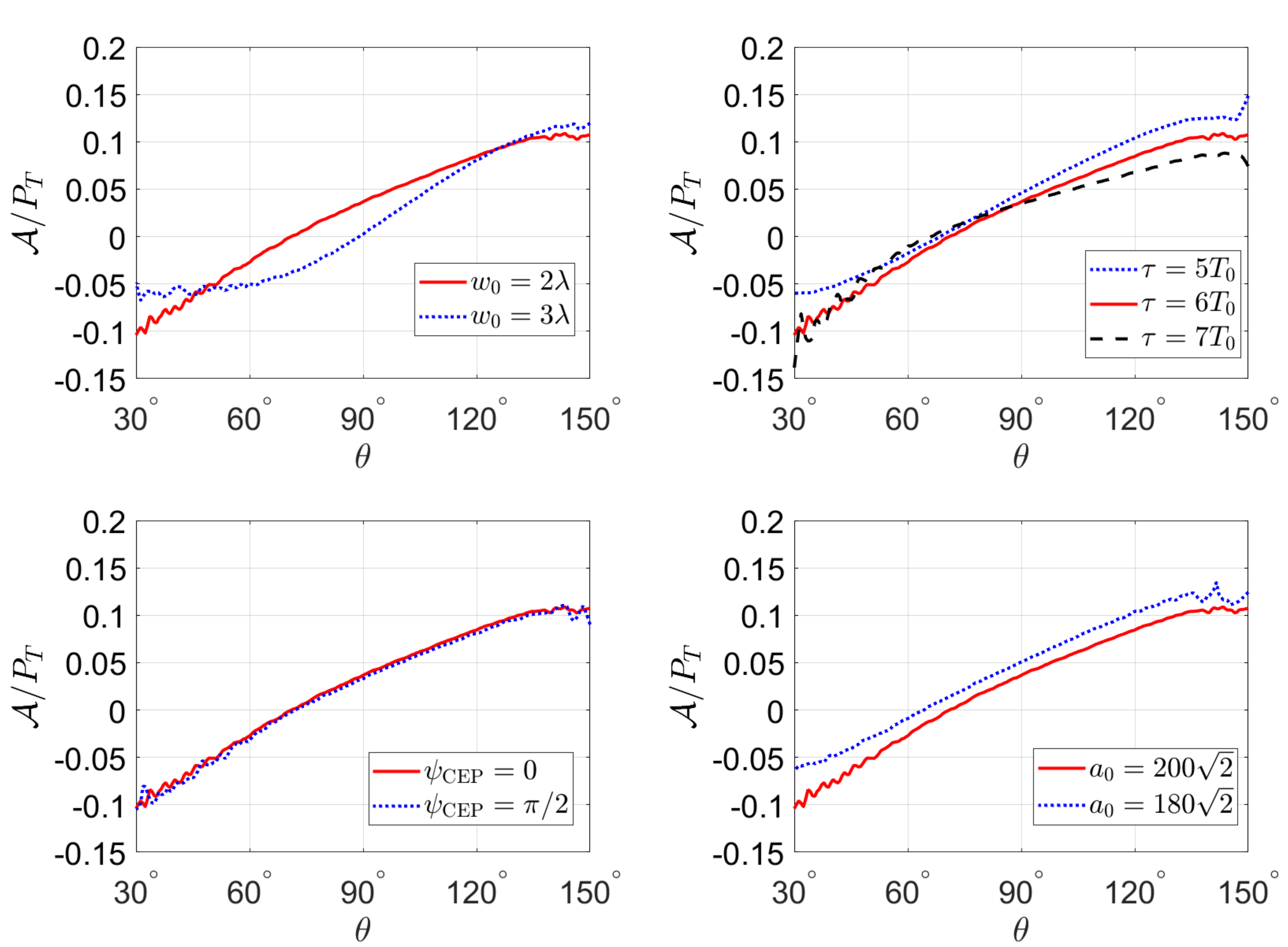}
  \begin{picture}(300,0)
  \put(28,175){\small (a)}
  \put(155,175){\small (b)}
    \put(28,84){\small (c)}
  \put(155,84){\small (d)}
 \end{picture}		
  \caption{Influence of fluctuations of laser pulse parameters, including focus radius $w_0$ (a), pulse duration $\tau$ (b), initial carrier-envelope phase $\psi_{\rm CEP}$ (c), and intensity $a_0$ (d), on the OLSE signature. } \label{flu_2}
  \end{figure}

To verify the feasibility of  the NRP observation,
 we investigate the influence of fluctuations of the laser and electron parameters on the value of the OLSE signal, see Figs.~\ref{flu_1} and  \ref{flu_2}.
   We vary the following parameters: initial electron kinetic energy $\varepsilon_0$, electron beam divergence $\Delta \theta_i/\theta_i$, the electron bunch length $L_e$, and the collision angle $\theta_{\rm collision}$; focus radius $w_0$, pulse duration $\tau$,  carrier-envelope phase $\psi_{\rm CEP}$ of the laser pulse, and intensity $a_0$ within the limits of the experimental feasibility  \cite{Danson2019}. It can be seen that the OLSE signature, i.e. a sizeable longitudinal polarization degree in the opposite direction to the driven laser, is robust under the influence of such fluctuations. Therefore, an averaging over the uncertainty of the variables would still allow for the detection of the considered effect under realistic conditions.

Let us give a rough estimation of the signal to noise ratio (SNR) considering the statistical noise for experimental detection. When selecting the scattered electrons with $\theta < 45.7^\circ$, we can get an average ${\cal A}/P_T$ of 0.075 which is one order of magnitude larger than the experimental detection precision. The relative fraction of electron density for these electrons is 0.23\%. Typically,  the charge of an electron bunch accelerated by laser-plasma interactions is  in a range of tens to hundreds pC, corresponding to the electron number of $10^8$. Then we get the SNR $\sim \sqrt{N} =478.6$ and the statistical uncertainty $\sim 1/\sqrt{N} =0.0021$. It could be demonstrated that the signal of the OLSE signature is strong enough for detection taking account of the statistical error.

\begin{figure}[b]
 	\includegraphics[width=1.0\linewidth]{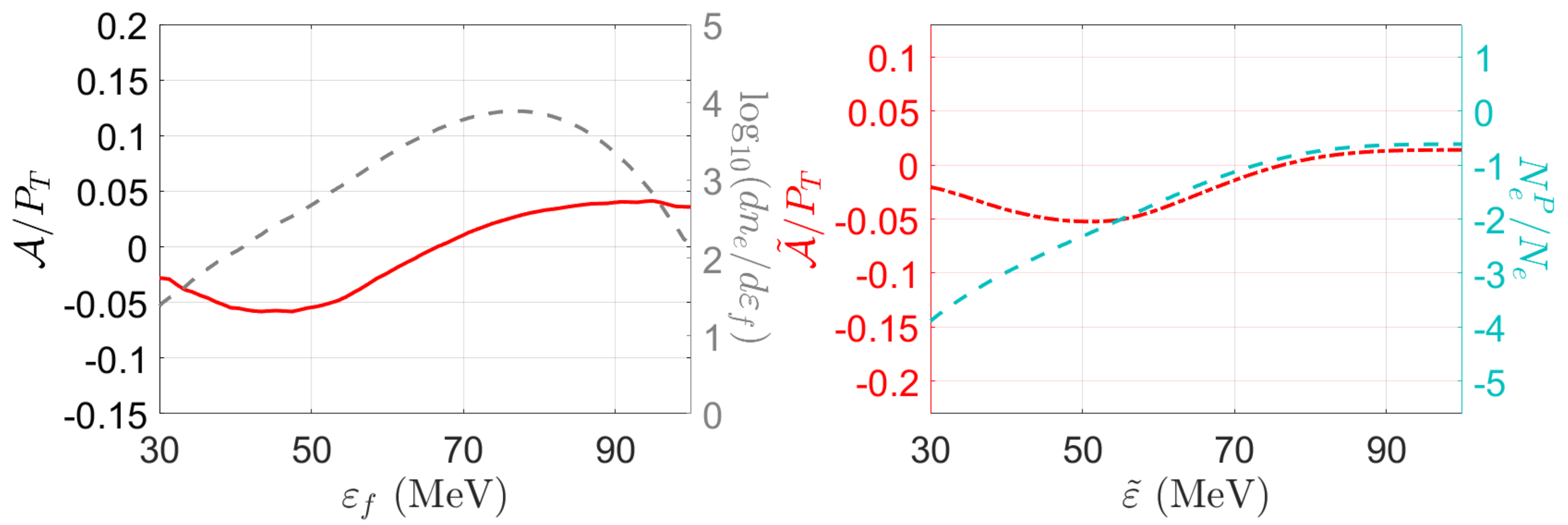}
  \begin{picture}(300,-30)
  \put(25,82){\small (a)}
  \put(148,82){\small(b)}
 \end{picture}
		  \caption{
  (a) $\mathcal{A}_{P_T}$ and number density of log$_{10}$(d$n_{e}$/d$\varepsilon_f$) (MeV$^{-1}$) vs $\varepsilon_f$; (b) Asymmetry parameter of selected electrons with energy $\varepsilon<\tilde{\varepsilon}$ (left axis) and their relative fraction (right axis)  vs $\tilde{\varepsilon}$. Here only forward electrons ($\theta<90^\circ$) are shown.} \label{energy}
\end{figure}
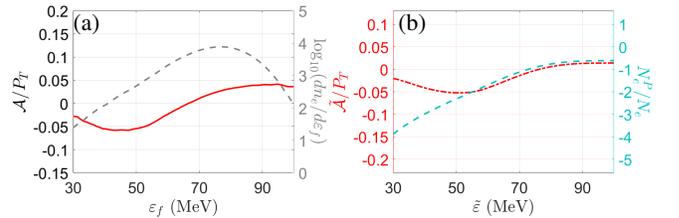
\begin{figure}      	
\includegraphics[width=1\linewidth]{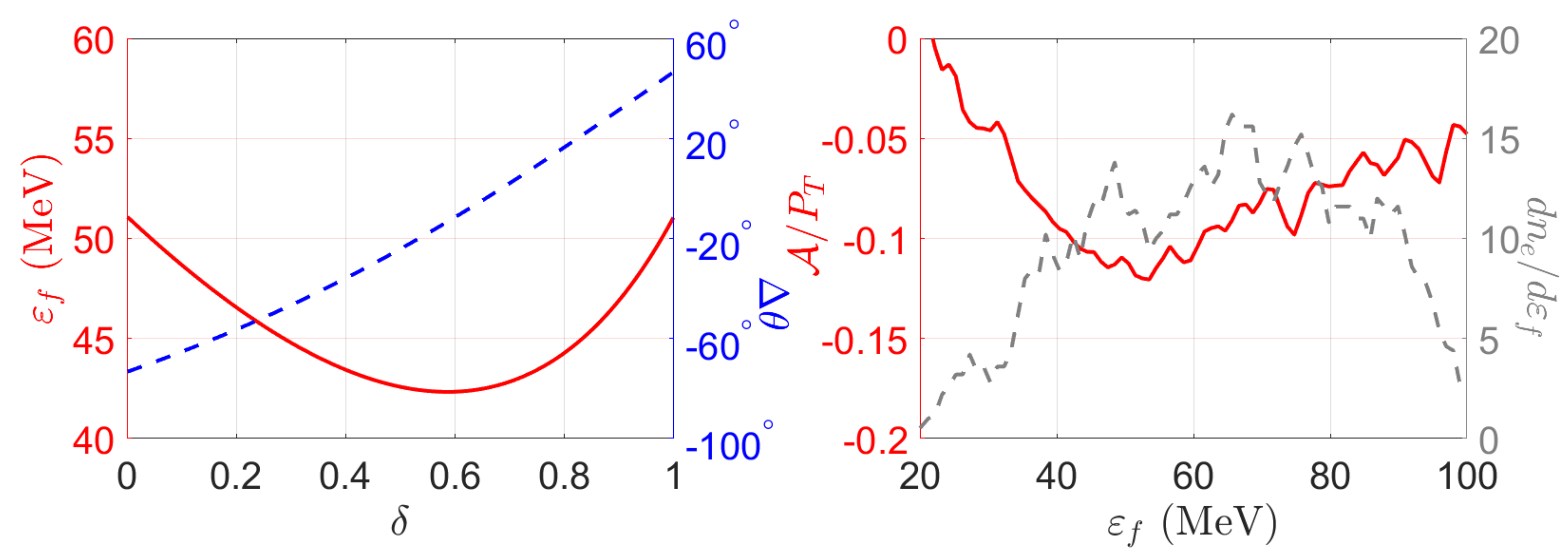}
\begin{picture}(300,0) 
  \put(25,82){\small(a)}   
   \put(150,82){\small(b)}
\end{picture}	
\caption{ (a) The final electron energy (red-solid) and deflection angle $\Delta\theta$ (blue-dashed) versus $\delta$ for a single emission, $\gamma=100$, $a_0=200$. Distributions vs final electron energy of $\varepsilon_f$ and $\theta$; (b) Distribution of $\mathcal{A}/P_T$ with respect to $\varepsilon_f$ for electrons with $\theta\in (0,40.25^\circ)$.} \label{postenergy}
\end{figure}
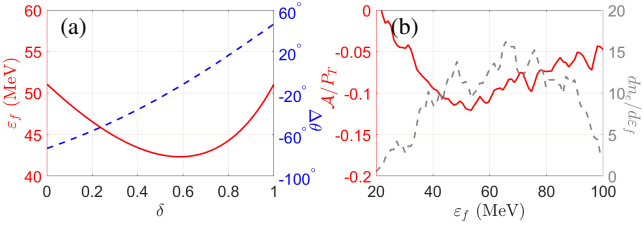

\subsection{The possibility of  NRP detection with the energy-selecting technique }

For forward scattered electrons, the  polarization dependence on the electron energy allows us to purify $\mathcal{A}$ with post-energy-selection [see Fig.~\ref{energy}].  One may naively expect that electrons with higher final energy, would have less radiative energy loss and should have larger $|\mathcal{A}|$ induced by NRP. However,  Fig.~\ref{energy}(a) indicates that the maximum of $|\mathcal{A}|$ is located at $\varepsilon=43.4$~MeV with $\mathcal{A}= -5.8\% P_T$.   This is because the radiation loss due to high energy photon emission is accompanied by  laser photons absorption, which leads to a nonlinear dependency of the electron final energy on the emitted photon energy. Therefore the energy-selecting technique is less sensitive  compared with the angle-selecting technique. 

Is it possible to detect NRP with energy post-selection? This question is analyzed in Fig.~\ref{postenergy}. As pointed out above, the energy of an electron is determined by two processes,  radiation energy loss due to high energy photon emissions and absorption of the laser photons, which accompanies the process. As a result the relation of $\varepsilon_f$ and  $\delta$ becomes nonlinear [see Fig. \ref{postenergy}(a)]. Therefore, electrons scattered to different $\theta$ could be closely located in the energy spectrum [Fig.~\ref{fig2}(a)], leading to a reduced sensitivity of $P_\parallel$ to the energy selection with respect to the angle one [Fig.~\ref{fig2}(b)]. The polarization degree could be further improved by combining post-angle-selection and post-energy-selection methods. For instance, $|\mathcal{A}|$ of electrons with $\theta \in (0, 40.25^\circ)$ could be enhanced to 11.3\% by selecting electrons with $\varepsilon\in[45\text{MeV},60 \text{MeV}]$, see Fig.\ref{postenergy}(b).


 \subsection{Influence of pair-production on the electron polarization}

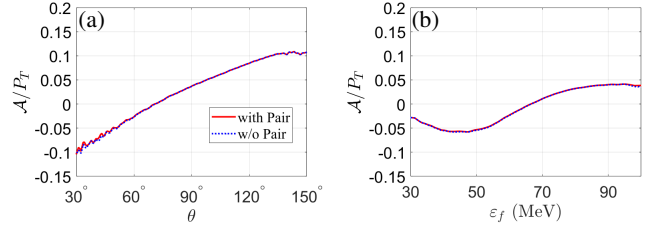
\begin{figure}[htb]
 	\includegraphics[width=1.0\linewidth]{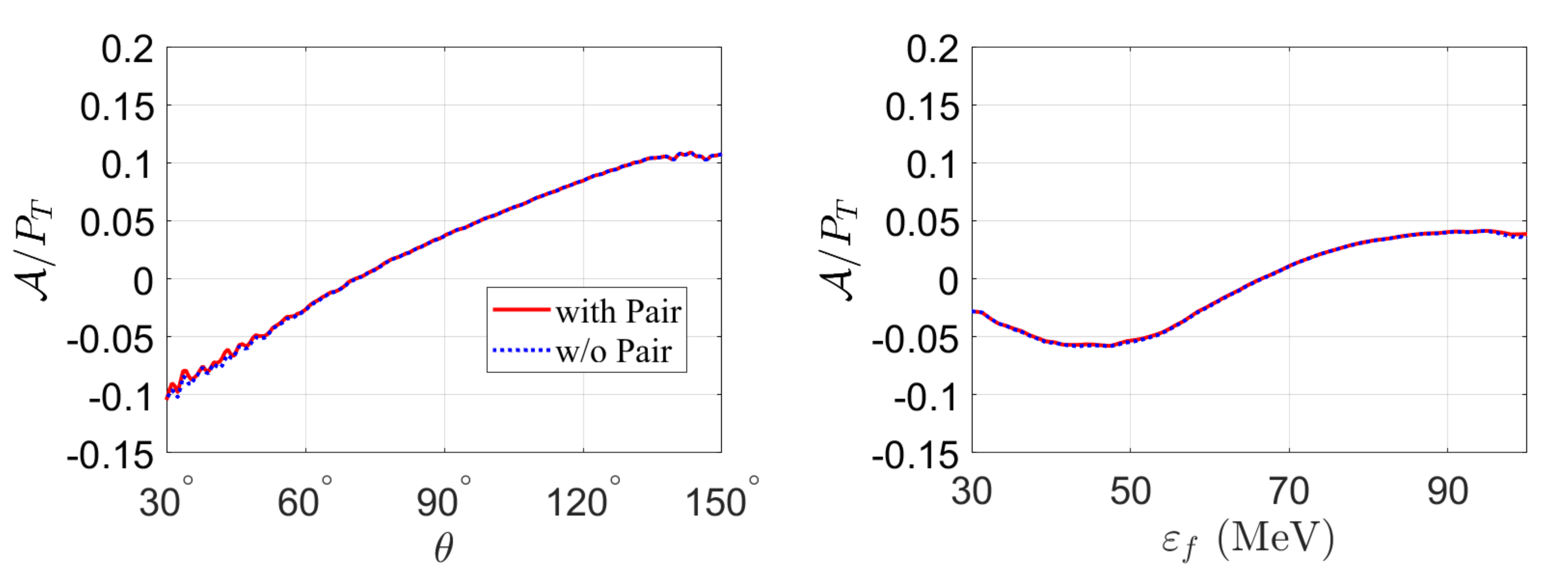}
  \begin{picture}(300,-30)
  \put(27,85){\small (a)}
  \put(153,85){\small(b)}

 \end{picture}
   \caption{Asymmetry parameter $\mathcal{A}$ vs  polar angle $\theta$ (a) and vs $\varepsilon_f$ (b). The red-solid curves indicate the numerical results with pair-production, while the blue-dotted curves correspond to those with pair-production effects removed artificially. In (b), only forward electrons ($\theta<90^\circ$) are shown.} \label{pair}
\end{figure}

We analyze the impact of the pair-production process for our parameters. The electrons created from the Breit-Wheeler pair-production process are included in the simulation result presented in Fig. \ref{pair}. The yield of the created electrons is only 0.11\% with respect to the initial electron amount. The ratio of the created $e^+e^-$ pairs to the emitted photons is $N_{e^+e^-}/N_\gamma \sim 1.7\times10^{-5}.$ Therefore, pair-production is negligible for the considered parameters and neglected in the above simulations.

\section{conclusion}\label{conclusion}

We have put forward a scheme to identify the distinct signature of the electron NRP, described by the contribution of the QED one-loop self-energy correction, in the electron and laser beam collision. 
Employing the radiation dominated reflection regime,  an experimentally accessible signal of electron helicity $P_\parallel \gtrsim 10\%$, opposite to the driving laser helicity for electrons collected within the specific small scattering angle will indicate the signature of the NRP effect. 
The signature is robust with respect to the laser and electron parameters and measurable with currently available experimental technology, providing a way for an experimental detection of the QED prediction on self-energy corrections via intense laser driven spin dynamics.

\section*{Acknowledgements}
Y-F.L. and Y-Y.C.
are supported by the National Natural Science Foundation of China (Grants No. 12222507, No. 12074262 and No. 12075187 ), the Strategic Priority Research Program of the Chinese Academy of Sciences (Grant No. XDA25031000), the National Key R\&D Program of China (Grant No. 2021YFA1601700), and the Shanghai Rising-Star Program.

\section*{APPENDIX A: The applied simulation method}

We employ a Monte Carlo method incorporating all the polarization effects in strong-QED processes  to simulate the interaction of a circularly polarized ultraintense laser pulse with an ultrarelativistic electron beam.  
In our simulations, the electron initial and final spins are resolved, while the
 polarization of the emitted photons is summed over. The energy of the emitted photon is determined by the spectral probability with the commonly used stochastic procedure \cite{Ridgers2014,Elkina2011,Gonoskov2015,CAIN}. The propagation direction of the emitted photon is along that of the emitting particle in the relativistic regime.

At each time step, the electron polarization vector $\mathbf{S}_i$ jumps to  $ \mathbf{S}_R$ after a photon emission. Here, the 3-vector $\mathbf{S}_R$ is the mean value of the electron final spin in its rest frame,  corresponding to
the electron average polarization state resulting from the scattering process itself \cite{Berestetskii1982}. When a photon emission does not take place at the given time step, nevertheless, the electron spin varies to the state of $\mathbf{S}_{NR}$ according to the no-emission probability \cite{CAIN}, which follows from the one-loop contribution to the electron mass operator \cite{Meuren2011,Ilderton2020,Torgrimsson_2021}.
Besides, the spin precession between emissions is governed by the Bargmann-Michel-Telegdi (BMT) equation  \cite{Bargmann1959}, where a field dependent anomalous magnetic moment is included as a result of the one-loop vertex correction \cite{Baier1972}.
Meanwhile, between quantum events of photon emissions, the electron dynamics in the laser field is described classically by the Lorentz equation.

In Ref.~\cite{Liyfei2020},  we have improved the Monte Carlo method
for simulating the strong-field QED processes incorporating the algorithm introduced in
\cite{Ridgers2014,Elkina2011,Gonoskov2015,CAIN}. In comparison with the existing spin-resolved laser-plasma (electron beam) interaction codes \cite{Chen2019,Liyf2019,Liyf2020,Wan2020,Seipt2019,Song2019}, the upgraded code allows for
the calculation of the three-dimensional polarization effects.

\subsubsection{Algorithm of event generation for photon emission}

In a time step $\Delta t$, the probability for an electron to emit a photon with an energy  $\omega_\gamma=\delta \varepsilon_i$ ($0<\delta<1$) is calculated with Eq.~(\ref{WR1}), i.e. $W_R(\delta)=\sum_{\pm S_f}\frac{{\rm d} W_R}{{\rm d}u {\rm d}t}\frac{{\rm d}u}{{\rm d}\delta} \Delta t$, where $\delta=r_1^3$, with $r_1$ being a random number in [0,1].  Another random number  $r_2\in [0,1]$ is used to determine if a photon is emitted: if $W_R(r_1)<r_2$,   the photon emission event is rejected; otherwise, emission of a photon with energy $\omega_\gamma=\delta \varepsilon_i$ occurs. Given the smallness of the emission angle $1/\gamma_e$ for the ultrarelativistic case, the photon is emitted along the electron velocity direction. Note that an approximation to the leading order in $1/\gamma_e$ is used throughout the paper. More detailed information on this method and its accuracy have been shown in Ref.\cite{Gonoskov2015}.
After the photon emission, the electron spin jumps to a mixed state $\mathbf{S}_R$ determined by Eq.(\ref{RP}).
Note that, one could also chose a pure spin state of $\pm \mathbf{S}_R/|\mathbf{S}_R|$ by using a random number \cite{Liyf2019,Liyf2020,Liyfei2020}, which  coincides with the former method except for a higher statistical fluctuation.  The former can be regarded as an averaged method  of the latter and the differences are negligible for a dense electron beam.
Meanwhile, if a photon emission event is rejected, the electron spin is also changed due to the no-emission probability, see Secs.~\ref{simpleman} and \ref{OLSE_calc}. 
The final polarization vector of the electron is ${\bf S}_{NR}$ via Eq.~(\ref{NP}).
Equation (\ref{dp_Baier}) and the discussion below it clarify why the momentum does not undergo a corresponding change in the non-radiative case.

\subsubsection{ Particle dynamics in the external laser field between photon emissions}

Between quantum events, the electron dynamics in the ultraintense laser field are described by the Lorentz equation
\begin{eqnarray}\label{L}
\frac{{\rm d}{\bf p}}{{\rm d}t}&=&e({\bf E+{\bm \beta}\times{\bf B}}).
\end{eqnarray}

 The spin precession is governed by the Thomas-Bargmann-Michel-Telegdi equation \cite{Bargmann1959}:
\begin{eqnarray}\label{BMT}
\frac{{\rm d}{\bf S}}{{\rm d}t}&=&\frac{e}{m}{\bf S}\times\left[-\left(\frac{g}{2}-1\right)\frac{\gamma}{\gamma+1}\left({\bm \beta}\cdot{\mathbf B}\right){\bm \beta}\right.\nonumber\\
&&\left.+\left(\frac{g}{2}-1+\frac{1}{\gamma}\right){\bf B}-\left(\frac{g}{2}-\frac{\gamma}{\gamma+1}\right){\bm \beta}\times{\bf E}\right],
\end{eqnarray}
where ${\bf E}$ and  ${\bf B}$ are the  laser electric and magnetic fields, respectively,
$g$ is the electron gyromagnetic factor:
$g\left(\chi_e\right)=2+2\mu\left(\chi_e\right)$ and $\mu\left(\chi_e\right)=\frac{\alpha}{\pi\chi_e}\int_{0}^{\infty}\frac{y}{\left(1+y\right)^3}{\bf L}_{\frac{1}{3}} \left(\frac{2y}{3\chi_e}\right){\rm d}y$ with ${\bf L}_{\frac{1}{3}} \left(z\right)=\int_{0}^{\infty}{\rm sin}\left[\frac{3z}{2}\left(x+\frac{x^3}{3}\right)\right]{\rm d}x$. As $\chi_e\ll1$, $g\approx2.00232$.

\subsubsection{Motion of the spin of an electron ensemble in a strong external field with inclusion of radiation effects}\label{Sec. C}

In our semiclassical Monte-Carlo simulation, the evolution of electron spin is determined by two parts: spin precession  between emissions via the BMT equation, and quantum spin variation due to the nature of radiation (including radiative polarization resulting from the emission of a real photon, and non-radiative variation originating from the emission and re-absorption of a virtual photon). The non-radiative polarization probability applied in our simulation can be derived from the interference of the OLSE and forward scattering diagrams [see Sec.\ref{OLSE_calc}], which additionally yields the spin evolution according to the modified spin precession due to the electron anomalous magnetic moment.
The OLSE contribution to the polarization is included in our simulation as NRP probability and the spin precession governed by BMT with the anomalous magnetic moment.
The faithfulness of our method is proved by the limiting procedure $\chi\ll 1$, when it yields the seminal
Baier equation (3.23) in \cite{Baier1972} for the spin evolution  in an external field.
Taking into account both spin precession and quantum spin variation, we can obtain the following equation for the motion of the spin of an ensemble of electrons in an external field,

\begin{eqnarray}\label{SKE}
\frac{d\mathbf{S}}{dt} &  =&\frac{e}{m}\left[\mathbf{S}\times\mathbf{F}\right]-\frac{\alpha m}{\sqrt{3}\pi\gamma}\int_0^\infty\frac{u^{2}du}{\left(1+u\right)^{3}}\left(\mathrm{K_{2/3}}\mathbf{S}\nonumber\right.\\
&&\left.+\left(\mathrm{IntK_{1/3}}-\mathrm{K_{2/3}}\right)\left(\mathbf{S}\cdot\bm{\beta}\right)\hat{{\bf e}}_v+\hat{{\bf e}}_2\mathrm{K_{1/3}}\right),\\
\mathbf{F}&=&\left[-\left(\frac{g}{2}-1\right)\frac{\gamma}{\gamma+1}\left({\bm \beta}\cdot{\mathbf B}\right){\bm \beta}\right.\left.+\left(\frac{g}{2}-1+\frac{1}{\gamma}\right){\mathbf B}\nonumber\right.\\\nonumber
&&\left.-\left(\frac{g}{2}-\frac{\gamma}{\gamma+1}\right){\bm \beta}\times{\mathbf E}\right]\nonumber.\\\nonumber
\end{eqnarray}
The first term corresponds to the BMT equation that governs the precession of the spin, taking into account  the electron anomalous magnetic moment, with $g-2$ being due to the QED radiative corrections.
This rotation term, which does not contain Planck's constant $\hbar$ explicitly (besides the implicit dependence via $g$),  can be obtained on the basis of a classical consideration and describes the precession of the magnetic moment.
The remaining terms (containing $\hbar$) are associated with the quantum spin variation, with and without photon emission. These terms are derived with Eqs.~(\ref{Svar})- (\ref{dSNR}). In contrast to the BMT term which does not change $|\mathbf{S}|$, the quantum terms lead to a damping effect of  $|\mathbf{S}|$.  Therefore, BMT and quantum spin variation between emissions have significantly different effects on spin evolution  and are both essential for a comprehensive description of spin dynamics in an external field.  

According to Eq.(\ref{SKE}), the accumulation of the longitudinal polarization of an ensemble of electrons is governed by the equation, see Ref.~\cite{Li2022}:
\begin{eqnarray}\label{P}
\frac{dS_\parallel}{dt}
&=&-\frac{e}{m}{\mathbf S_\perp} \cdot \left[\left(\frac{g}{2}-1\right) {{\bm { \beta}}} \times {\bf B} + \left(\frac{g \beta}{2}-\frac{1}{\beta}\right) {\bf E}\right]\nonumber\\
&& -\frac{\alpha m}{\sqrt{3}\pi\gamma} S_\parallel\int_0^\infty\frac{u^{2}du}{\left(1+u\right)^{3}}\mathrm{IntK_{1/3}}(u').
\end{eqnarray}

\section*{APPENDIX B: The electromagnetic fields of the laser pulses}

In this work, we employ tightly-focused laser pulses with a Gaussian temporal profile, which propagate along $-z$ direction as a scattering laser beam. The spatial distribution of the electromagnetic fields takes into account up to $\epsilon_0^3$-order of the nonparaxial solution, where $\epsilon_0=w_0/z_r$,  $w_0$ is the laser focal radius, $z_r=k_0w_0^2/2$ the Rayleigh length,  $k_0=2\pi/\lambda_0$ the laser wave vector, and $\lambda_0$ the laser wavelength. The expressions of the electromagnetic fields of the linearly-polarized (LP) (along $x$ axis) laser pulse are as follows \cite{Yousef2002}:
\begin{eqnarray}
E_x^{(L)} &=& -i E^{(L)}\left[1+\epsilon_0^2\left(f^2\widetilde{x}^2-\frac{f^3\rho^4}{4} \right) \right],\nonumber\\
E_y^{(L)} &=& -i E^{(L)} \epsilon_0^2 f^2 \widetilde{x}\widetilde{y},\nonumber\\
E_z^{(L)} &=& -E^{(L)}\left[\epsilon_0 f \widetilde{x} + \epsilon_0^3 \widetilde{x} \left(-\frac{f^2}{2}+f^3\rho^2-\frac{f^4\rho^4}{4}\right) \right],\nonumber\\
B_x^{(L)} &=& 0, \\
B_y^{(L)} &=& i E^{(L)} \left[1+\epsilon_0^2\left(\frac{f^2\rho^2}{2}-\frac{f^3\rho^4}{4} \right)\right],\nonumber\\
B_z^{(L)} &=& E^{(L)}\left[\epsilon_0 f \widetilde{y} + \epsilon_0^3 \widetilde{y} \left(\frac{f^2}{2}+\frac{f^3\rho^2}{2}-\frac{f^4\rho^4}{4}\right) \right],\nonumber\\
E^{(L)} &=& E_0 F_n f e^{-f\rho^2} e^{i\left(\psi+\psi_{\rm{CEP}}\right)} e^{-\frac{t^2}{\tau^2}}, \nonumber
\end{eqnarray}
where $\tau$ is the laser pulse duration, $E_0$ the amplitude of the laser fields with normalization factor $F_n=i$ in order to provide in the focal spot  $E_0=\sqrt{|E_x^{(L)}|^2+|E_y^{(L)}|^2+|E_z^{(L)}|^2}$, with  scaled coordinates
\begin{eqnarray}
\widetilde{x}&=&\frac{x}{w_0}, \quad \widetilde{y}=\frac{y}{w_0},\quad \widetilde{z}=\frac{z}{z_r}, \quad \rho^2=\widetilde{x}^2+\widetilde{y}^2,\\
f&=&\frac{i}{\widetilde{z}+i},
\end{eqnarray}
the laser field phase $\psi=\omega_0 t+k_0z$, and the carrier-envelope phase $\psi_{\rm{CEP}}$.\\

The circularly-polarized (CP) laser pulses can be assumed to be the combination of two orthogonal LP laser pulses, polarized along $x$ and $y$ directions, respectively, with a $\pi/2$ phase delay:
\begin{eqnarray}
E_x&=&E_x^{(1)}+E_x^{(2)},\quad E_y=E_y^{(1)}+E_y^{(2)},\quad E_z=E_z^{(1)}+E_z^{(2)},\nonumber\\
B_x&=&B_x^{(1)}+B_x^{(2)},\quad B_y=B_y^{(1)}+B_y^{(2)},\quad B_z=B_z^{(1)}+B_z^{(2)},\nonumber
\end{eqnarray}
where
\begin{eqnarray}
E_x^{(1)} &=& -i E^{(1)}\left[1+\epsilon_0^2\left(f^2\widetilde{x}^2-\frac{f^3\rho^4}{4} \right) \right],\nonumber\\
E_y^{(1)} &=& -i E^{(1)} \epsilon_0^2 f^2 \widetilde{x}\widetilde{y},\nonumber\\
E_z^{(1)} &=& -E^{(1)}\left[\epsilon_0 f \widetilde{x} + \epsilon_0^3 \widetilde{x} \left(-\frac{f^2}{2}+f^3\rho^2-\frac{f^4\rho^4}{4}\right) \right],\nonumber\\
B_x^{(1)} &=& 0,\nonumber\\
B_y^{(1)} &=& i E^{(1)} \left[1+\epsilon_0^2\left(\frac{f^2\rho^2}{2}-\frac{f^3\rho^4}{4} \right)\right],\nonumber\\
B_z^{(1)} &=& E^{(1)}\left[\epsilon_0 f \widetilde{y} + \epsilon_0^3 \widetilde{y} \left(\frac{f^2}{2}+\frac{f^3\rho^2}{2}-\frac{f^4\rho^4}{4}\right) \right];\nonumber
\end{eqnarray}
\begin{eqnarray}
E_x^{(2)} &=& i E^{(2)} \epsilon_0^2 f^2 \widetilde{x}\widetilde{y},\nonumber\\
E_y^{(2)} &=& i E^{(2)}\left[1+\epsilon_0^2\left(f^2\widetilde{y}^2-\frac{f^3\rho^4}{4} \right) \right],\nonumber\\
E_z^{(2)} &=& E^{(2)}\left[\epsilon_0 f \widetilde{y} + \epsilon_0^3 \widetilde{y} \left(-\frac{f^2}{2}+f^3\rho^2-\frac{f^4\rho^4}{4}\right) \right],\nonumber\\
B_x^{(2)} &=& i E^{(2)} \left[1+\epsilon_0^2\left(\frac{f^2\rho^2}{2}-\frac{f^3\rho^4}{4} \right)\right],\nonumber\\
B_y^{(2)} &=& 0,\nonumber\\
B_z^{(2)} &=& E^{(2)}\left[\epsilon_0 f \widetilde{x} + \epsilon_0^3 \widetilde{x} \left(\frac{f^2}{2}+\frac{f^3\rho^2}{2}-\frac{f^4\rho^4}{4}\right) \right];\nonumber\\
E^{(1)} &=& E_0 F_n f e^{-f\rho^2} e^{i\left(\psi+\psi_{\rm{CEP}}\right)} e^{-\frac{t^2}{\tau^2}},\nonumber\\
E^{(2)} &=& E_0 F_n f e^{-f\rho^2} e^{i\left(\psi+\pi/2+\psi_{\rm{CEP}}\right)} e^{-\frac{t^2}{\tau^2}},
\end{eqnarray}
and the superscripts $^{(1)}$ and $^{(2)}$ denote the laser pulses linearly-polarized along $x$ and $y$ directions, respectively.

\section*{APPENDIX C: Description of electron dynamics in the reflection regime}

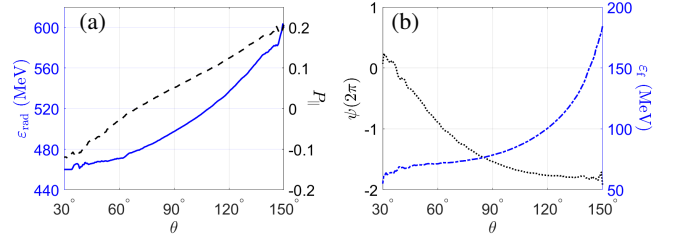
\begin{figure}[b]
  	\includegraphics[width=1\linewidth]{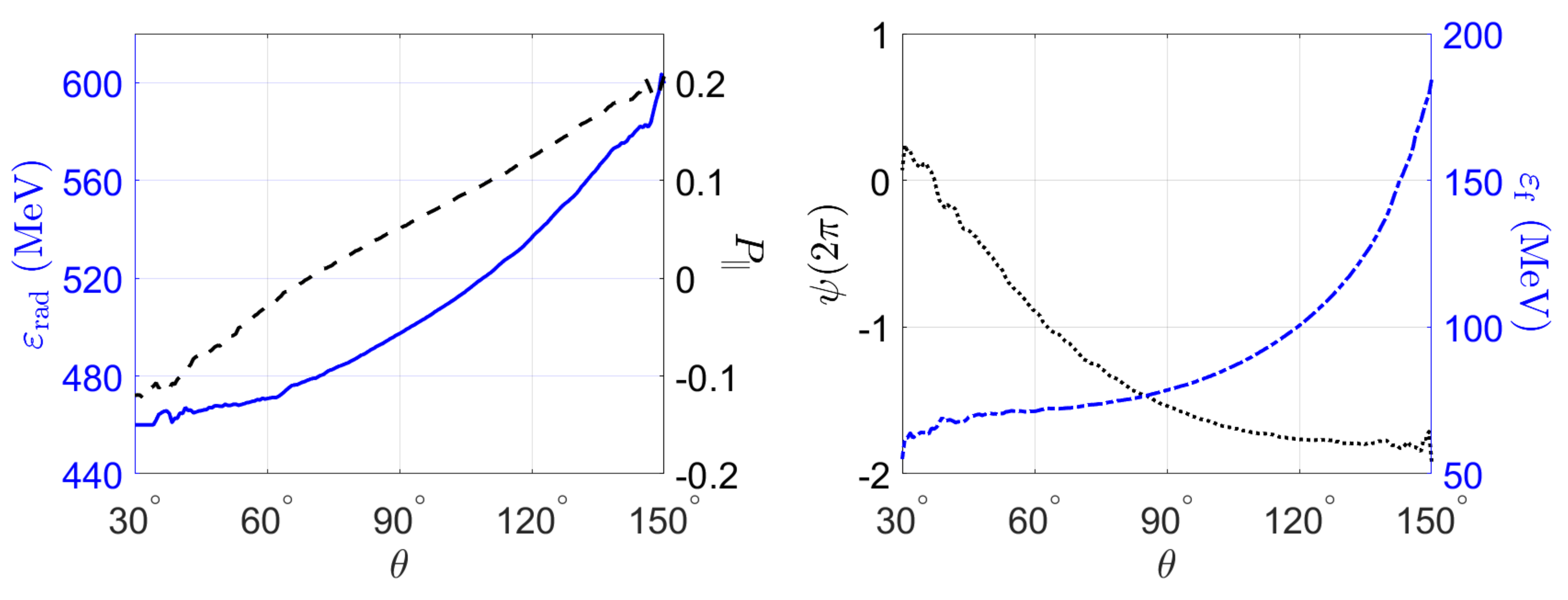}
  \begin{picture}(300,0)
  \put(27,91){\small (a)}
  \put(145,91){\small(b)}
 \end{picture}		
  \caption{(a): Average radiation energy loss and $P_\parallel$ vs $\theta$. (b): Average phase (time)  $\psi=\omega_0 t+k_0 z$ of multiple photon-emission events and final electron energies vs $\theta$.  Here, $\psi=0$ refers to the peak of the laser pulse. } \label{dynamics}
  \end{figure}

The electron dynamics in the reflection regime is analyzed in Fig.~\ref{dynamics}. We plotted the distribution curves of average radiation energy loss, average phase (time)  $\psi=\omega_0 t+k_0 z$ of multiple photon-emission events, and average final energy of electrons with respect to $\theta$. It can be seen that electrons with $P_\parallel>0$ experience more stochastic radiation loss (also radiative polarization effect), while those with $P_\parallel<0$ experience less stochastic radiation loss, i.e. more nonradiative polarization.  Electrons undergoing more radiation energy loss at the beginning of the laser pulse tend to be reflected and accelerated to higher energies by the laser field. Electrons experiencing less radiation energy loss would penetrate deeper into the laser pulse, and emit photon with larger phase $\psi$ [see Fig.~\ref{dynamics}(b)]. When radiation energy loss is small enough, the OLSE could surpass SFE, making the negative helicity associated with OLSE dominates over positive helicity associated with SFE.

We provide a rough estimation of the deflection angle corresponding to the average emitted photon energy $\overline{\delta}_c$ relevant for RP, which  can be estimated as
\begin{eqnarray}
\Delta \theta_c \sim \frac{\left[2-\bar{\delta}_{c}(1+{\rm cos}\theta_{\gamma})\right]\bar{\delta}_{c}{\rm sin}\theta_{\gamma}}{2+\bar{\delta}_{c}(\bar{\delta}_{c}-2)(1+{\rm cos}\theta_{\gamma})}
\end{eqnarray}.

We estimate the final electron momentum and energy. For simplicity, we use a circularly polarized monochromatic plane-wave  for estimating the final electron momentum and energy. The  momentum and energy of electrons in the laser can be represent as \cite{Piazza2012}:
\begin{eqnarray}
{\bf p}_\perp(\psi) &=& {\bf p}_{\perp,0}-e{\bf A}(\psi),\\ \nonumber
p_\parallel(\psi) &=& p_{\parallel,0}+\frac{e^2[{\bf A}(\psi)]^2}{2 \Lambda},\\\nonumber
\varepsilon(\psi) &=& \varepsilon_{\parallel,0}+\frac{e^2[{\bf A}(\psi)]^2}{2 \Lambda},
\end{eqnarray}
where ${\bf A}(\psi)= \hat{\bf i} A_0 \cos(\psi)+\hat{\bf j} A_0 \sin(\psi)$ is the potential of the wave with $A_0=m a_0/e$, $\Lambda=k\cdot p$, $p_{\perp,0}=0$ and $p_{z,0}\approx \gamma$. When a photon with angular frequency $\omega_\gamma$ is emitted, then we get the emission angle
\begin{eqnarray}
{\rm sin}\theta_\gamma  &=&  \frac{p_\perp}{\sqrt{{p}_\perp^2+p_\parallel^2}}\approx\frac{p_\perp(\psi)}{\varepsilon(\psi)}\approx\frac{a_0}{\gamma+a_0^2/4\gamma}.
\end{eqnarray}
The final electron energy and momentum can be obtained via the energy-momentum conservation:
\begin{equation}\label{conserve}
p+n k=p'+k',
\end{equation}
with $p,p'$ being the electron four-momentum before and after the interaction, and $k,k'$ the incident and outgoing photon four-momentum, and $n$ the number of photons absorbed. Here we can get:
\begin{equation}
n=\frac{k'p}{kp-k'k}\approx \frac{\omega_\gamma \varepsilon_0(1-{\rm cos}\theta_\gamma)}{2\omega_0\varepsilon_0-\omega_\gamma\omega_0(1+{\rm cos}\theta_\gamma)}.
\end{equation}
The final   transverse momentum and energy of electrons out of laser can be obtained from Eq. (\ref{conserve}),
\begin{eqnarray}
p_\perp^f &=& -\omega_\gamma {\rm sin} \theta_\gamma,\\\nonumber
\varepsilon_f&=&\varepsilon_0-\omega_\gamma+n\omega_0\approx \varepsilon_0 \frac{2+\delta(\delta-2)(1+{\rm cos}\theta_\gamma)}{2-\delta(1+{\rm cos}\theta_\gamma)},
\end{eqnarray}
and the  electron deflection angle:
\begin{equation}
\theta_e=\sin^{-1} \frac{p_\perp^f}{\varepsilon_f}\approx  \sin^{-1} \left\{\frac{[2-\delta(1+{\rm cos}\theta_\gamma)]\delta{\rm sin}\theta_\gamma}{2+\delta(\delta-2)(1+{\rm cos}\theta_\gamma)}\right\}.
\end{equation}

\bibliography{reference.bib}

\end{document}